\documentstyle[prb,aps,multicol,epsfig]{revtex}
\begin{document}
\title{Superradiance of low density Frenkel excitons \\
in a crystal slab of three-level atoms: Quantum interference
effect}
\author{G. R. Jin\cite{PresentAdd} , P. Zhang, Yu-xi
Liu$^{\dag}$, C. P. Sun\cite{email,www}}
\address{Institute of Theoretical Physics, Academia Sinica, P.O. Box 2735, Beijing 100080, China}
\address{$^{\dag}$The Graduate University for Advanced Studies (SOKEN-DAI),
         Hayama, Kanagawa, 240-0193, Japan}
\date{\today}
\maketitle
\begin{abstract}
We systematically study the fluorescence of low density Frenkel
excitons in a crystal slab containing $N_T$ V-type three-level
atoms. Based on symmetric quasi-spin realization of SU(3) in large
$N$ limit, the two-mode exciton operators are invoked to depict
various collective excitations of the collection of these V-type
atoms starting from their ground state. By making use of the
rotating wave approximation, the light intensity of radiation for
the single lattice layer is investigated in detail. As a quantum
coherence effect, the quantum beat phenomenon is discussed in
detail for different initial excitonic states. We also test the
above results analytically without the consideration of the
rotating wave approximation and the self-interaction of radiance
field is also included. 
\end{abstract}
\begin{multicols}{2}
\section{Introduction}
It is well known that the fluorescence of an exciton exhibits
super-radiant character due to the appearance of  macroscopic
transition dipole moment of the exciton
\cite{Hopfield,Agrannovich1,Hanamura,Itoh,Andreani1,Konester1,Konester2,Bjork1,Bjork2,Agrannovich2,Beer,Fidder}.
However this  collective feature of exciton radiance depends upon
dimensionality of crystal. In infinite bulk crystals, the exciton
does not radiate because this exciton is dressed with a photon
which has the same wave vector due to the transitional symmetry of
the total system. As a result, a stable polariton is formed
\cite{Hopfield}. In the case of lower dimensional systems, one can
show that the exciton decays superradiantly due to breakdown of
the translational symmetry. For example the decay rates of exciton
are of the order $(\lambda /a)\gamma $ for one-dimensional (1D)
crystals and $(\lambda /a)^{2}\gamma $ for 2D crystals
\cite{Agrannovich1}, with $\gamma $ being the radiative decay rate
for an isolated atom, $a$ the lattice constant, and $\lambda $ the
light wavelength.

The enhanced factor $(\lambda /a)^{D}$ in the radiative width of
exciton is now regarded commonly as one of evidences of
``superradiance" \cite{Agrannovich2}, which has been demonstrated
in experiments for Frenkel excitons \cite{Aaviksoo} and Wannier
excitons \cite{Deveaud}. Since the enhanced radiant effect can
also appear in the usual superradiance of an ensemble of atoms
\cite{sf,sf1}, how can we distinguish the different features
between the fluorescence of the excitons and that of the atomic
ensemble? It is known that, in the fully population-inverted atom
systems, the atoms radiate independently with each other in the
initial stage. The back-action of the emitted photon to the atoms
results in the correlation among atoms. Consequently, the atoms
become cooperative and thus the fluorescence from the atomic
ensemble will show different statistical properties in the initial
and final stages. On the other hand, the exciton fluorescence
exhibits identical statistical character during the whole process
\cite{Cao}. Physically, this is because the initial dipole moments
of the atoms are spatially random in an atomic ensemble, but in a
semiconductor crystal, the dipole moment of exciton presents a
macroscopic effect even at initial moment. So the optical
properties of multidimensional quantum-confined semiconductor
structures (MQCS), such as quantum wells, quantum wires and
quantum dots, have their own specific features in physical
processes.

The optical properties of the MQCS in a semiconductor microcavity
(SMC) have attracted more and more attention in the past decades
\cite{Yamamoto}. The SMC with high-reflectivity dielectric mirrors
leads to the realization of the strong coupling between radiation
and matter. Moreover, since the optical mode structure of the SMC
may alter around the MQCS, many new phenomena, such as tailoring
the spontaneous-radiation rate and pattern \cite{Sp1,Sp2,Sp3}, the
coupled exciton-photon mode splitting in a SMC \cite{SMC,Liu0},
have been demonstrated. The resonant interaction between excitons
and a single-mode cavity field and the corresponding detuning
effect were further investigated \cite{Liu1,Liu2}.

Most of these former works mentioned above dealed only with the
two-level lattice atom case, however the three-level atom case may
be very useful to implement quantum information encoding and
processing \cite{lukin0,lukin1,Flei0,Flei1,zol0}. Over the last
few years, the cavity QED with the collective excitations of
ensembles of three-level atoms has attracted much attention for
quantum computing implementations. In this case, many atoms are
entangled through their interaction with the common cavity field.
To maintain quantum coherence in this quantum information
processing \cite{Sun}, it is important to reach the so-called
strong coupling regime where the single-photon coherent coupling
$g_0\gg \gamma\text{, }\gamma_{\text{cav}}$, the atomic and cavity
dissipation (decoherence) rates respectively. It is the symmetric
collective excitation that can reach the strong coupling regime
without requiring a high finesse cavity as $g_0\propto \sqrt{N}$,
with the total number of atoms $N$ and here we need to consider
the quantum decoherence  induced by  the spatial inhomogeneousness
of couplings \cite{sun-you,liyong}. In fact, the phenomena of
superfluorescence or superradiance \cite{{n3,n4,tom}} constitute
another example of collective state dynamics. Recent experimental
success clearly demonstrates the power of such an atomic ensemble
based on the system for entangling macroscopic objects
\cite{polzik}.

In the present paper, we study the fluorescence of low density
excitons in a crystal slab containing V-type three-level atoms.
The purpose of this paper is to investigate the quantum
interference effect \cite{three-level} in the time evolution of
light intensity. In sec. II, considering that the existence of the
two-level atomic exciton is mathematically associated with the
infinite dimensional reducible representation of SU(2) Lie algebra
\cite{Sun2}, we can rationally define the exciton operators for
the three-level case associated with SU(3) algebra, which contains
various SU(2) subalgebras. With this conception, both the free
part and the interaction part of Hamiltonian can be written down
in terms of the introduced two-mode exciton operators, which can
be described by bosonic operators in the low-density limit. In
sec. III, with rotating wave approximation (RWA) for the
interaction $(e/mc)\mathbf{P}\cdot\mathbf{A}$, the coupled
equations for the exciton-light field system are solved with the
help of Weisskopf-Wigner approach (WWA) \cite{Meystre,Scully}. In
sec. IV, the explicit expressions of the electric field operators
are derived for the monolayer case. The light intensity as well as
the first- and second-order degree of quantum coherence are
calculated to show certain new features of exciton fluorescence in
a three-level crystal slab. We discuss the phenomenon of quantum
beat in the time evolution of the light intensity for various
initial exciton states. In sec. V, we consider the roles of both
the non-RWA terms and the self-interaction term of photon. This
consideration avoids ``unphysical" roots of the characteristic
equation when the non-perturbation approach given in Ref.
\cite{Cao} is used. The light intensity is calculated to compare
with the results in sec. IV. The conclusions are presented in Sec.
VI with some remarks.

\section{two mode exciton system with SU(3) structure}
We consider a plane crystal slab with a simple cubic structure
which contains a stack of $N$ identical layers. V-type three-level
atoms, as shown in Fig. 1, occupy $N_T$ lattice sites, where
$N_T=N_{L}N$ and $N_{L}$ is the total lattice sites within each
layer. The wave vectors of the excitons and light fields are all
assumed perpendicular to the slab. We restrict ourselves to
investigate only the low density excitons region.
\begin{figure}[t]
\begin{center}
\epsfxsize=5.5cm\epsffile{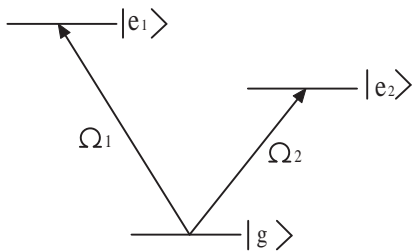}\\[0pt]
\epsfxsize=5.5cm\epsffile{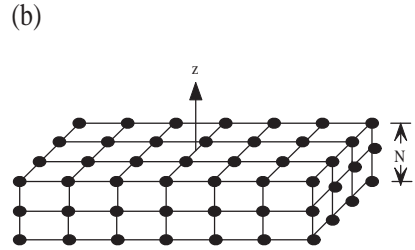}
\end{center}
\par
\vskip 0.3cm \caption{(a) Energy structure of a V-type three-level
atom. $\Omega_{1}$is the transition frequency of
$|g\rangle\leftrightarrow|e_{1}\rangle$, and $\Omega_{2}$ is that
of $|g\rangle\leftrightarrow |e_{2}\rangle$. The direct transition
between the two upper states is forbidden. (b) A plane crystal
slab with simple cubic structure containing a stack of $N$
identical layers with $N_L$ sites within each layer.}
\end{figure}

The interaction $\frac e{mc}\mathbf{p}\cdot \mathbf{A}$ between
the radiation field and the multi-atom system is written in the
second quantization form
\begin{eqnarray}
\hat{H}_1&=&\hbar \sum_{q;l,j}g_1(q)\left| e_1\right\rangle
_{lj}\left\langle g\right| \left[\hat{a}_q+\hat{a}_{-q}^{\dagger
}\right] e^{iqla} \nonumber \\
&&+\hbar \sum_{q;l,j}g_2(q)\left|
e_2\right\rangle _{lj}\left\langle g\right|
\left[\hat{a}_q+\hat{a}_{-q}^{\dagger }\right]
e^{iqla}+\mathrm{H.c.},
\end{eqnarray}
where $\hat{a}_q$ and $\hat{a}_q^{\dagger }$ are the annihilation
and creation operators of the photon with wave vector $q$ along
$z$, respectively, $j$ and $l$ denote the $j$th lattice site in
the $l$th layer. $g_1(q)=\sqrt{\frac{2\pi \Omega _1}{V\hbar
|q|c}}d_1$, $g_2(q)=\sqrt{\frac{2\pi \Omega _2}{V\hbar |q|c }}d_2$
are the effective atom-photon coupling constants of $\left|
g\right\rangle \leftrightarrow \left| e_1\right\rangle $ and
$\left| g\right\rangle \leftrightarrow \left| e_2\right\rangle$
atomic transitions with transition frequencies $\Omega_1$ and
$\Omega_2$, respectively. The corresponding transition dipole
moments are represented by ${\bf d}_1$ and ${\bf d}_2$. We assume
that ${\bf d}_1$ and ${\bf d}_2$ are parallel and lying in the
slab plane.

Introduce the collective operators for the $l$th layer as
\cite{Cao}
\begin{eqnarray}
\hat{A}^{(l)\dagger}&=&\frac{1}{\sqrt{N_{L}}}\sum_{j}\left|
e_{1}\right\rangle _{lj}\left\langle g\right|
,\hat{A}^{(l)}=\frac{1}{\sqrt{ N_{L}}}\sum_{j}\left|
g\right\rangle_{lj}\left\langle e_{1}\right| ,
\nonumber \\
\hat{B}^{(l)\dagger} &=&\frac{1}{\sqrt{N_{L}}}\sum_{j}\left|
e_{2}\right\rangle _{lj}\left\langle g\right|
,\hat{B}^{(l)}=\frac{1}{\sqrt{ N_{L}}}\sum_{j}\left|
g\right\rangle _{lj}\left\langle e_{2}\right|.
\end{eqnarray}
It is easy to prove that, two sets of the operators
\begin{eqnarray}
E_1^{\left(l\right)}&=&\sum_j\left| e_1\right\rangle
_{lj}\left\langle g\right| ,F_1^{\left(l\right)
}=E_1^{\left(l\right)\dagger },
\end{eqnarray}
and
\begin{eqnarray}
F_2^{\left(l\right)} &=&\sum_j\left| e_2\right\rangle
_{lj}\left\langle g\right| ,E_2^{\left(l\right)
}=F_2^{\left(l\right)\dagger},
\end{eqnarray}
just generate two SU(2) algebras not commuting with each other.
This means the four collective operators $E_1^{\left(l\right)}$,
$E_2^{\left(l\right)}$, $F_1^{\left(l\right)}$ and
$F_2^{\left(l\right)}$ do not span a product algebra
SU(2)$\otimes$SU(2). A straightforward calculation in Appendix A
checks that they satisfy SU(3) algebra. Actually, the above four
collective operators define a spinor realization of SU(3) of
$N_L+1$ dimensions. Furthermore, the unique number $N_L$, atom
number, determines the dimensions $N_L+1$ of representations of
the two SU(2) subalgebras. Since we understood the single-mode
exciton in terms of the large $N_L$ limit of representations of
SU(2), it is easy to prove for SU(3) case that, in a very large
$N_L$ and the low density excitation region, the above defined
collective operators $\hat{A}^{(l)}$ and $\hat{B}^{(l)}$ become
the bosonic ones obeying the commutation relation
\begin{equation}
\left[\hat{A}^{(l)},\hat{A}^{(l')\dagger}\right] =\delta
_{l,l^{\prime
}},\left[\hat{B}^{(l)},\hat{B}^{(l')\dagger}\right]=\delta
_{l,l^{\prime }},
\end{equation}
where the ideal bosonic approximation is equivalent to the neglect
of the phase space filling effect and the exciton-exciton
interaction \cite{Hanamura0}.

In terms of these collective operators, the two-mode Frenkel
exciton operators with wave vectors $k=\frac{2\pi m}{Na}$
($m=-\frac 12(N-1),-\frac 12(N-3),...,\frac 12(N-1)$), are just
the discrete Fourier transformations for them
\begin{eqnarray}
\hat{A}_k &=&\frac 1{\sqrt{N}}\sum_le^{-ikla}\hat{A}^{(l)},\nonumber \\
\hat{B}_k &=&\frac 1{\sqrt{N}}\sum_le^{-ikla}\hat{B}^{(l)}.
\end{eqnarray}
In fact their conjugates $A_{k}^{\dagger }$ and $B_{k}^{\dagger }$
are just the generators for the quasi-spin wave states
\begin{eqnarray}
|A_k\rangle = \hat{A}_{k}^{\dagger }|g_1,g_2,...,g_{N_T}\rangle,
\nonumber \\
|B_k \rangle =\hat{B}_{k}^{\dagger}|g_1,g_2,...,g_{N_T}\rangle.
\end{eqnarray}
Since the operators $\hat{A}_k$ and $\hat{B}_k$ commute with each
other for very large $N_L$ and low excitation, they form a two
independent boson system. Then we obtain the interaction
Hamiltonian for the the two-mode Frenkel exciton system coupled to
a quantized electromagnetic field
\begin{eqnarray}
\hat{H}_1 &=&\hbar \sum_{q,k}G_1(q)O(k+q)\left[
\hat{A}_k+\hat{A}_{-k}^{\dagger }\right] \left[
\hat{a}_q+\hat{a}_{-q}^{\dagger }\right]\nonumber\\
&+&\hbar \sum_{q,k}G_2(q)O(k+q)\left[
\hat{B}_k+\hat{B}_{-k}^{\dagger }\right] \left[
\hat{a}_q+\hat{a}_{-q}^{\dagger }\right],
\end{eqnarray}
where the coupling constants between the photons and excitons take
the following form
\begin{eqnarray}
&&G_{1}(q)=\sqrt{N_T}g_1(q)=\sqrt{N_T\frac{2\pi \Omega
_1^2}{V\hbar \omega_q}}d_1,
\nonumber \\
&&G_{2}(q)=\sqrt{N_T}g_2(q)=\sqrt{N_T\frac{2\pi \Omega
_2^2}{V\hbar \omega_q}}d_2,
\end{eqnarray}
here $\omega_q=|q|c$. The wave-vector matching factor \cite{Cao}
in Eq. (8) is
\begin{equation}
O(k+q)=\frac 1N\sum_le^{i(k+q)la}=\frac 1N\frac{\sin
(\frac{k+q}2Na)}{\sin (\frac{k+q}2a)},
\end{equation}
which is real and equal to 1 for $k+q=0$, and $O(k+q)<1$, for
$k+q\neq 0$.

\section{Equations of motion with rotating wave approximation}
In this section, we consider the rotating wave approximation (RWA)
to deal with the interaction $\frac e{mc}\mathbf{p}\cdot
\mathbf{A}$ between the radiation field and the multi-atom system.
After introducing the two-mode exciton operators in Eq. (6), the
interaction Hamiltonian between the excitons and photons with RWA
is obtained as
\begin{eqnarray}
\hat{H}_{RWA}&=&\hbar\sum_{q,k}G_1(q)O(k+q)\left[\hat{A}_k
\hat{a}_{-q}^{\dagger }+\hat{A}_{-k}^{\dagger}
\hat{a}_q\right]\nonumber\\
&&+\hbar \sum_{q,k}G_2(q)O(k+q)\left[\hat{B}_k
\hat{a}_{-q}^{\dagger }+\hat{B}_{-k}^{\dagger }\hat{a}_q \right],
\end{eqnarray}
where we have neglected the higher frequency (non-resonant) terms:
$\hat{A}^{\dagger }_{-k}\hat{a}^{\dagger }_{-q}$,
$\hat{A}_{k}\hat{a}_{q}$, $\hat{B}^{\dagger }_{-k}\hat{a}^{\dagger
}_{-q}$, and $\hat{B}_{k}\hat{a}_{q}$.

The Heisenberg equations for the exciton and photon operators are
\begin{eqnarray}
i\frac \partial {\partial t}\hat{A}_k &=&\Omega _1\hat{A}_k+
\sum_qG_1(q)O(q-k) \hat{a}_q , \\
i\frac \partial {\partial t}\hat{B}_k &=&\Omega _2\hat{B}_k+
\sum_qG_2(q)O(q-k) \hat{a}_q , \\
i\frac \partial {\partial t}\hat{a}_q &=&\omega_q\hat{a}_q+G_1(q)
\sum_kO(k-q) \hat{A}_k  \nonumber \\
&&+G_2(q)\sum_kO(k-q)\hat{B}_k. \label{eq:aqRWA}
\end{eqnarray}
Taking the following transform
\begin{eqnarray}
\hat{A}_k \rightarrow \widetilde{A}_k=\hat{A}_k e^{i \Omega_1 t},
\hat{B}_k \rightarrow \widetilde{B}_k=\hat{B}_k e^{i \Omega_2 t},
\end{eqnarray}
to remove the fast varying factors, we obtain the formal equation
for the exciton operator
\begin{eqnarray}
\frac \partial {\partial t}\widetilde{A}_k(t)&=&
-i\sum_qG_1(q)O(q-k)e^{-i(\omega_q-\Omega_1)t}\hat{a}_q(0) \nonumber \\
&&-\sum_{q,k^{\prime}}G_1(q)G_2(q)O(q-k)O(k^{\prime}-q)  \nonumber \\
&&\times\int_0^{t}\widetilde{B}_{k^{\prime}}(t^{\prime})
e^{-i(\omega_q-\Omega_1)t}e^{i(\omega_q-\Omega_2)t^{\prime}}dt^{\prime}
\nonumber \\
&&-\sum_{q,k^{\prime}}G_1^2(q)O(q-k)O(k^{\prime}-q)\nonumber \\
&&\times\int_0^{t}\widetilde{A}_{k^{\prime}}(t^{\prime})
e^{-i(\omega_q-\Omega_1)(t-t^{\prime})}dt^{\prime}.\label{eq:dak}
\end{eqnarray}
Here, the first term proportional to $\hat{a}_q(0)$ is the
so-called quantum noise term. The second term in the above
equation corresponds to a multi-photon processes (MPP) including
stimulated emission and absorption effects. Its contribution can
be ignored since it is a higher order term from the standpoint of
perturbation \cite{Meystre}. The last term in Eq. (\ref{eq:dak})
can be solved by using the Weisskopf-Wigner approximation (WWA),
i.e., assuming that $\widetilde{A}_{k^{\prime}}(t^{\prime})$
varies sufficiently slowly so that it can be factorized outside
the integral. The remaining part of the time integral of the last
term in Eq. (\ref{eq:dak}) can be evaluated and get a Dirac
$\delta$ function with variable $(\omega_q-\Omega_1)$ and a
principal part ${\mathcal
P}\left[\frac{i}{\omega_q-\Omega_1}\right]$ term, which
contributes a frequency shift (Lamb shift).

The equation of motion for $\widetilde{B}_k(t)$ can also be
obtained in the similar way. By using the WWA, and neglecting the
MMP, we can solve the equations for both of the two-mode excitons
to obtain $\widetilde{A}_k(t)$ and $\widetilde{B}_k(t)$.
Substituting them into Eq. (\ref{eq:aqRWA}), the explicit
expression for $\hat{a}_q(t)$ in terms of $\hat{a}_q(0)$, $\hat{A}
_k(0)$ and $\hat{B}_k(0)$ can be obtained. Finally we can get the
positive-frequency part of the electric field
\begin{eqnarray}
\hat{E}^{(+)}(z,t)&=&i\sum_q \sqrt{\frac{2\pi\hbar
\omega_q}{V}}\hat{a}
_q(t)e^{iqz}  \nonumber \\
&=&i\int_{-\infty}^{\infty}dq \sqrt{\frac{\hbar \omega_q L}{2\pi
A}} \hat{a} _q(t)e^{iqz},
\end{eqnarray}
where we have taken the photon normalization volume $V$ to be
$LA$, with $A$ being the area of the crystal slab, and assumed
that the slab is located at the middle of the volume. When $L$ is
sufficiently large, the sum over $q$ has been replaced by an
integral: $\sum_{q}...\rightarrow
\frac{L}{2\pi}\int_{-\infty}^{\infty}dq...$.

For an arbitrary initial state
$|\phi\rangle=|\phi_{ex}\rangle\otimes|\phi_{L}\rangle$ of the
total system, the light intensity radiated from the two-mode
exciton system is defined as
\begin{equation}
I(z,t)=\frac{c}{2\pi}\langle \phi|:\hat{E}^{(-)}(z,t)\hat{E}
^{(+)}(z,t):|\phi\rangle,
\end{equation}
where the symbol ``: ... :" means the normal product according to
the exciton operators. $z$ is the position of a detector relative
to the crystal slab. The first-order degree of coherence of
fluorescence is given by
\begin{equation}
g^{(1)}(z;t,t+\tau)=\frac{\frac{c}{2\pi}\langle \phi|:\hat{E}^{(-)}(z,t)
\hat{E}^{(+)}(z,t+\tau):|\phi\rangle}{\sqrt{I(z,t)I(z,t+\tau)}}.
\end{equation}
Similarly, one can also define the second-order degree of
coherence as
\end{multicols}
\begin{widetext}
\begin{eqnarray}
g^{(2)}(z;t,t+\tau)=\frac{\frac{c^2}{4\pi^2}\langle
\phi|:\hat{E}^{(-)}(z,t)
\hat{E}^{(-)}(z,t+\tau)\hat{E}^{(+)}(z,t+\tau)\hat{E}^{(+)}(z,t):
|\phi\rangle}{I(z,t)I(z,t+\tau)}.
\end{eqnarray}
\end{widetext}

\section{Quantum interference effect in light intensity}
\begin{multicols}{2}
In this section, the features of exciton fluorescence in the case
of monolayer will be studied in detail. From the definition of the
wave vector of excitons, when $N=1$, it takes only one value $k=0$
and the wave-vector matching factor $O(q-k)=O(k^{\prime}-q)\equiv
1$. Thus with the help of WWA, we get the equations of motion for
exciton modes $A_0$ and $B_0$ as
\begin{eqnarray}
\frac \partial {\partial
t}\widetilde{A}_0(t)&=&-i\sum_qG_1(q)e^{-i(\omega_q-\Omega_1)t}
\hat{a}_q(0)-\frac{\eta_1}{2}\widetilde{A}_0(t),
\label{eq:eofm1}\\
\frac\partial {\partial
t}\widetilde{B}_0(t)&=&-i\sum_qG_2(q)e^{-i(\omega_q-\Omega_2)t}
\hat{a}_q(0)-\frac{\eta_2}{2}\widetilde{B}_0(t), \label{eq:eofm2}
\end{eqnarray}
in which,
\begin{eqnarray}
\eta_1&=&\frac{4\pi\Omega_1d_1^2}{\hbar a^2 c}=3\left(\frac{\pi
\lambda_1^2}{a^2}\right)\gamma_1,\label{eq:eta1} \\
\eta_2&=&\frac{4\pi\Omega_2d_2^2}{\hbar a^2 c}=3\left(\frac{\pi
\lambda_2^2}{ a^2}\right)\gamma_2,\label{eq:eta2}
\end{eqnarray}
where, $\gamma_1=\frac{4\Omega_1^3d_1^2}{3\hbar c^3}$ is the
radiative decay rate of an isolated atom from $|e_1\rangle$ to
$|g\rangle$, and $\gamma_2= \frac{4\Omega_2^3d_2^2}{3\hbar c^3}$
is that of $|e_2\rangle$ to $|g\rangle$ , respectively.
$\lambda_j=\frac{c}{\Omega_j}$, for $j=1\text{, }2$, denote the
reduced wavelengthes. We find that, because of the implementation
of WWA and the ignorance of multi-photon scattering processes
(MPP) in the derivation of  Eqs. (\ref{eq:eofm1}) and
(\ref{eq:eofm2}), the two-mode excitons decay independently in an
exponential rule. Even though our treatment seems to be bold, we
can get a useful result that the decay rates of excitons in two
dimensional crystal slab are proportional to the enhanced factor
$(\lambda/a)^2$. In section V, we will consider a more complex
case, in which a non-perturbation approach is used to restudy the
radiative decay rates of the two-mode excitons without WWA, and
the effects of MPP will be also investigated there.

Substituting the solutions of Eqs. (\ref{eq:eofm1}) and
(\ref{eq:eofm2}) into Eq. (\ref{eq:aqRWA}), we get the explicit
expression of the photon annihilation operator $\hat{a}_{q}(t)$
for the monolayer case:
\begin{eqnarray}
\hat{a}_q(t)&=&u_q(t)\hat{A}_{0}(0)
+v_q(t)\hat{B}_{0}(0)\nonumber\\
&&+e^{-i\omega_qt}\hat{a}_q(0)+\sum_{q'}w_{q,q'}(t)\hat{a}_{q'}(0),
\end{eqnarray}
where
\begin{eqnarray}
u_{q}(t)&=&G_1(q)\frac{e^{-i\omega_{q}t}-e^{-
\eta_{1}t/2}e^{-i\Omega_{01}t}}
{\omega_q-\Omega_{01}+i\eta_1/2},
\end{eqnarray}
\begin{eqnarray}
v_{q}(t)&=&G_2(q)\frac{e^{-i\omega_{q}t}-e^{-\eta_{2}t/2}e^{-i
\Omega_{02}t}} {\omega_q-\Omega_{02}+i\eta_2/2},
\end{eqnarray}
\begin{eqnarray}
w_{q,q'}(t)&=&\frac{G_1(q)G_1(q') e^{-i\omega _qt}}{\omega
_{q'}-\Omega_{01}+i\eta_1 /2}\left[ \frac{ e^{i(\omega
_q-\Omega_{01} )t}e^{-\eta_1 t/2}-1}{\omega_q-\Omega_{01}+i\eta_1
/2}\right.\nonumber\\
&&\left.-\frac{e^{i(\omega_q-\omega_{q'})t}-1}{\omega_q-\omega
_{q'}}\right]+\text{same with $1\rightarrow 2$},
\end{eqnarray}
in which $\Omega_{0j}=\Omega_j+\Delta_j$, for $j=1\text{, }2$, are
the renormalized physical frequencies, and $\Delta_j$ are the lamb
frequency shifts
\begin{equation}
\Delta_j=-{\mathcal P}\int_0^\infty
d\omega_q\frac{\rho(\omega_q)\left|
G_j(q)\right|^2}{\omega_q-\Omega_j}.
\end{equation}
Here ``${\mathcal P}$'' denotes for the Cauthy principle part, and
$\rho(\omega_q)$ is the density of states of the radiation field.

Therefore, one can obtain the positive frequency part of the
electric field operator
\begin{eqnarray}
\hat{E}^{(+)}(z,t)&=&\hat{E}^{(+)}_0(z,t)+
\sqrt{\frac{\pi\hbar\Omega_1 \eta_1}{4Ac}}F_{A}(z,t)\hat{A
}_{0}(0)\nonumber\\
&&+\sqrt{\frac{\pi\hbar\Omega_2
\eta_2}{4Ac}}F_{B}(z,t)\hat{B}_{0}(0)\label{eq:EzRWA},
\end{eqnarray}
where
\begin{eqnarray}
\hat{E}^{(+)}_0(z,t)&=&i\sum_q\sqrt{\frac{2\pi
\hbar\omega_q}{V}}\left[\hat{a}_q(0)e^{-i\omega_q
t}\right.\nonumber\\
&&\left.+\sum_{q'}w_{q,q'}(t)\hat{a}_{q'}(0)\right]e^{iqz}.
\end{eqnarray}
The first term in $\hat{E}^{(+)}_0(z,t)$ is just the free varying
photon field. The second term is proportional to the square of
coupling constants. As mentioned in Eq. (17), the sum over the
wave vector $q$ can be replaced by an integral, thus two
time-dependent functions $F_A(z,t)$ and $F_B(z,t)$ in Eq. (30) are
\begin{eqnarray}
F_{A}(z,t)&=&\frac{i}{\pi}\int_0^{\infty}d\omega_{q}2
\cos\left(\omega_{q}\frac{z}{c}\right)\frac{e^{-i\omega_{q}t}-e^{-i\omega_{1}t}}
{\omega_{q}-\omega_{1}},
\end{eqnarray}
\begin{eqnarray}
F_{B}(z,t)&=&\frac{i}{\pi}\int_0^{\infty}d\omega_{q}2
\cos\left(\omega_{q}\frac{z}{c}\right)\frac{e^{-i\omega_{q}t}-e^{-i\omega_{2}t}}
{\omega_{q}-\omega_{2}}.
\end{eqnarray}
Here $\omega_{1}=\Omega_{01}-i\eta_1/2$, $\omega_{2}=\Omega_{02}
-i\eta_2/2$. In the Ref. \cite{Cao}, similar expressions were
presented. However, the authors of Ref. \cite {Cao} did not
calculate the integrals over $d\omega_{q}$, which may result in
the light intensity going into the spacelike region. In this
paper, we will further calculate the integral over $d\omega _{q}$
\cite{Meystre,Chow} to give an explicit expression of
$\hat{E}^{(+)}(z,t)$.
\begin{eqnarray}
F_{A}(z,t) &=&\frac{1}{\pi }\int_{0}^{t}dt^{\prime }e^{-\eta
_{1}t^{\prime }/2}e^{-i\Omega _{01}t}\int_{0}^{\infty }d\omega
_{q}2\cos\left(\omega _{q}\frac{z }{c}\right)\nonumber\\
&&\times e^{i(\omega _{q}-\Omega_{01})(t'-t)}.
\end{eqnarray}
Replacing the integral variable $\omega _{q}$ by $\omega
_{q}-\Omega_{01}$, we have
\begin{eqnarray}
F_{A}(z,t)&=&\frac{1}{\pi }\int_{0}^{t}dt^{\prime }e^{-\eta
_{1}t^{\prime }/2}e^{-i\Omega_{01}t}\int_{-\Omega_{01}}^{\infty
}d\omega _{q}\nonumber\\
&&\times 2\cos \left[\frac{\left(\omega _{q}+\Omega
_{01}\right)z}{c}\right] e^{i\omega _{q}(t^{\prime }-t)}.
\end{eqnarray}
We assume that $\Omega_{01}$ is much larger than all other
quantities of the dimension of frequency \cite{Liu0}, so the lower
limit of the integration can be extended to $-\infty$. Therefore,
the above integral over $d\omega _{q}$ can be approximated by two
$\delta $ functions with variables $(t^{\prime}-t+\frac{z}{c})$
and $(t^{\prime }-t-\frac{z}{c})$. In the region $z>0$ outside the
layer, performing the integral over $ t^{\prime }$, we get
\begin{eqnarray}
F_{A}(z,t) &=&2e^{-\eta _{1}(t-\frac{z}{c})/2}e^{-i\Omega
_{01}(t-\frac{z}{c}
)}\Theta \left(t-\frac{z}{c}\right)  \nonumber \\
&=&2e^{-i\omega _{1}(t-\frac{z}{c})}\Theta
\left(t-\frac{z}{c}\right),\label{eq:fa0}
\end{eqnarray}
where $\Theta$ is a Heaviside function.  Similarly we can also
obtain
\begin{equation}
F_{B}(z,t)=2e^{-i\omega_{2}(t-\frac{z}{c})}\Theta
\left(t-\frac{z}{c}\right).\label{eq:fb0}
\end{equation}
From Eqs. (\ref{eq:fa0}) and (\ref{eq:fb0}), one can find that
there is no advanced {\it propagator} proportional to
$e^{-i(...)(t+\frac{z}{c})}$ appeared in the positive frequency
part of the electric field operator $\hat{E}^{(+)}(z,t)$. So that
the system is causal and well-behaved \cite{Meystre}.

It should be pointed out that $F_{A}(z,t)$ and $F_{B}(z,t)$ in Eq.
(\ref{eq:fa0}) and Eq. (\ref{eq:fb0}) show that the electric field
of the fluorescence emitted from the two-mode excitons is a
temporal damped plane wave, but does not decay with $z$. A
realistic electric field, however, should decrease with $z$, which
happens, especially when a light propagates from a medium with low
refractive index to that of higher one. Besides, a realistic
electric field should not simply be a plane wave. However, the
theoretical calculations with this subtle consideration will be
very complicated and do not provide us with necessary physics in
general. Our ideal assumption of a plane wave propagating in $z$
may work well for longer wave length radiation near the crystal.

The vacuum state of the exciton may be defined as
\begin{equation}
\hat{A}_0(0)\left| 0\right\rangle=\hat{B}_0(0)\left|
0\right\rangle =0,
\end{equation}
which represents that there is no exciton contained in the crystal
slab. Then number states for the two-mode excitons are
\begin{eqnarray}
&&\left| n\right\rangle _A=\frac1{\sqrt{n!}}\left[
\hat{A}_0^{\dagger }(0)
\right] ^n\left| 0\right\rangle, \nonumber\\
&&\left| m\right\rangle _B=\frac 1{\sqrt{m!}}\left[
\hat{B}_0^{\dagger }(0) \right]^m\left| 0\right\rangle.
\end{eqnarray}
The coherent states can be formally defined as eigenvectors of
exciton annihilation operators
\begin{eqnarray}
&&\hat{A}_0(0)\left| \alpha \right\rangle _A =\alpha \left| \alpha
\right\rangle _A,  \nonumber \\
&&\hat{B}_0(0)\left| \beta \right\rangle _B =\beta \left| \beta
\right\rangle _B.
\end{eqnarray}
It is noticed that this definition can work well only in the low
excitation case. This is because the generic expansion of a
coherent state concerns the Fock states with higher excitations.
However, the coherent state with smaller average exciton numbers
can still approximately describe the quantum coherence natures of
the exciton systems. Actually, we can understand the Fock state or
the coherent state in terms of the single atomic states (for the
details please see the appendix A). In this sense, the initial
state of Frenkel exciton can be written according to the initial
preparations of the single atomic states. A simplest illustration
is that the vacuum state of Frenkel exciton is just the state
formed by all atoms in the ground state. According to the appendix
A, the Fock state of Frenkel exciton is a symmetrized many-atom
state with certain atoms in the excited state. Especially, the
coherent state of Frenkel exciton is an atomic ( SU(2)) coherent
state.

The fluorescence of the two-mode excitons will be studied in the
following of this section. If one take the initial state of total
system as $|\phi\rangle=|\phi_{ex}\rangle\otimes|\{0\}\rangle$,
i.e., all modes of the light field are initially in vacuum state,
$\hat{E}^{(+)}_0(z,t)$ in the positive frequency part of the field
operator of Eq. (\ref{eq:EzRWA}) gives zero contribution.
Therefore, in the following calculations, we can neglect safely
the ``quantum noise", the terms proportional to $\hat{a}_q(0)$,
and investigate only the influence of various initial exciton
states on the fluorescence of the low density excitations in the
crystal slab.

We find that the light field emitted from the two-mode exciton
system contains two eigenmodes: $\Omega_{01}$ and $\Omega_{02}$.
These two modes of the light field can give an interference if
there are non-diagonal terms in the light intensity. To see this,
we need calculate the light intensity firstly. Substituting Eq.
(\ref{eq:EzRWA}) and its Hermitian conjugate into Eq. (18), the
light intensity is
\end{multicols}
\begin{widetext}
\begin{eqnarray}
I(z,t)&=&\frac{1}{8} \hbar\Omega _1\frac{\eta _1}A\left|
F_A(z,t)\right|^2\left\langle\hat{A}_0^{\dag}(0)
\hat{A}_0(0)\right\rangle+\frac{1}{8}\hbar \Omega _2\frac{\eta
_2}A\left| F_B(z,t)\right|^2\left\langle\hat{B}_0^{\dag}(0)
\hat{B}_0(0)\right\rangle\nonumber\\
&&+\frac{1}{8}\hbar \Omega_3\frac{\eta_3}A\left[
F_A^{*}(z,t)F_B(z,t)\left\langle\hat{A}_0^{\dag}(0)
\hat{B}_0(0)\right\rangle+C.c.\right], \label{eq:intensity}
\end{eqnarray}
\end{widetext}
\begin{multicols}{2}
\noindent where $\langle...\rangle$ denotes ensemble average,
$\Omega _3=\sqrt{\Omega _1 \Omega _2}$, and $\eta _3=\sqrt{\eta
_1\eta _2}$. The first two terms in Eq. (\ref{eq:intensity})
represent the average intensities of the two eigenmodes:
$\Omega_{01}$ and $\Omega_{02}$, respectively. The last term
proportional to $\frac{1}{8}\hbar
\Omega_3\frac{\eta_3}A\left[...\right]$, however, gives a
contribution of interference of the light fields.

When the exciton system is initially in a separable state
$\rho(0)=\rho_A\otimes \rho_B$ and both $\rho_A$ and $\rho_B$ can
be diagonal in Fock representation, there are only two terms in
the light intensity, which represent the contributions from the
two-mode excitons, respectively. There is no non-diagonal term in
the light intensity due to
$\left\langle\hat{A}_0^{\dag}(0)\hat{B}_0(0)\right\rangle=0$. The
light intensity is
\begin{eqnarray}
I(z,t)&=&\frac{1}{8} \hbar\Omega _1\frac{\eta _1}A\left\langle
n\right\rangle_A
\left[4e^{-\eta_1(t-z/c)}\right.\nonumber\\
&&\left.+4\chi
e^{-\eta_2(t-z/c)}\right]\Theta(t-z/c),\label{eq:intenFock}
\end{eqnarray}
where $\chi=\chi_0\left\langle m\right\rangle_B/\left\langle
n\right\rangle_A$, $\chi_0=\Omega_2\eta_2/\Omega_1\eta_1$, and
$\left\langle n\right\rangle_A$ ($\left\langle m\right\rangle_B$)
is the initial mean number of $A_0$($B_0$)-mode excitons. Note
that $\chi_0$ is only determined by the intrinsic properties of
the 2D sample. For a fixed $\chi_0$, $\chi$ can be used to
describe the degree of unsymmetrically excitation within the
crystal slab initially. Eq. (\ref{eq:intenFock}) is plotted in
Fig. 2, and our results show that $I(z,t)$ increases abruptly
after the propagation time $t=2\pi/\Omega_{1}$ then decays
exponentially and no interference pattern appears in this case.
When $\langle m\rangle_B=0$, i.e., the $B_0$-mode excitons are
initially in vacuum state, as shown in Fig. 2 (the solid line),
our result will go back to the two-level crystal atoms case
\cite{Cao} due to the using of RWA and neglecting MPP in the
derivation of Eq. (\ref{eq:eofm1}) and Eq. (\ref{eq:eofm2}), i.e.,
the mode of exciton being initially in the vacuum state gives no
contribution to the dynamics of the total system. Besides, from
the dot line ($\left\langle m\right\rangle_B/\left\langle
n\right\rangle_A=1$) and dashed-dot line ($\left\langle
m\right\rangle_B/\left\langle n\right\rangle_A=5$) of Fig. 2, we
find that, with the increase of $\chi$, the amplitude of the light
intensity becomes higher, which shows clearly that the amplitude
of the light intensity is the sum of the contributions of the two
eigenmodes of the radiation emitted from the two-mode excitons.
\begin{figure}[hbtp]
\begin{center}
\epsfxsize=7.0cm\epsffile{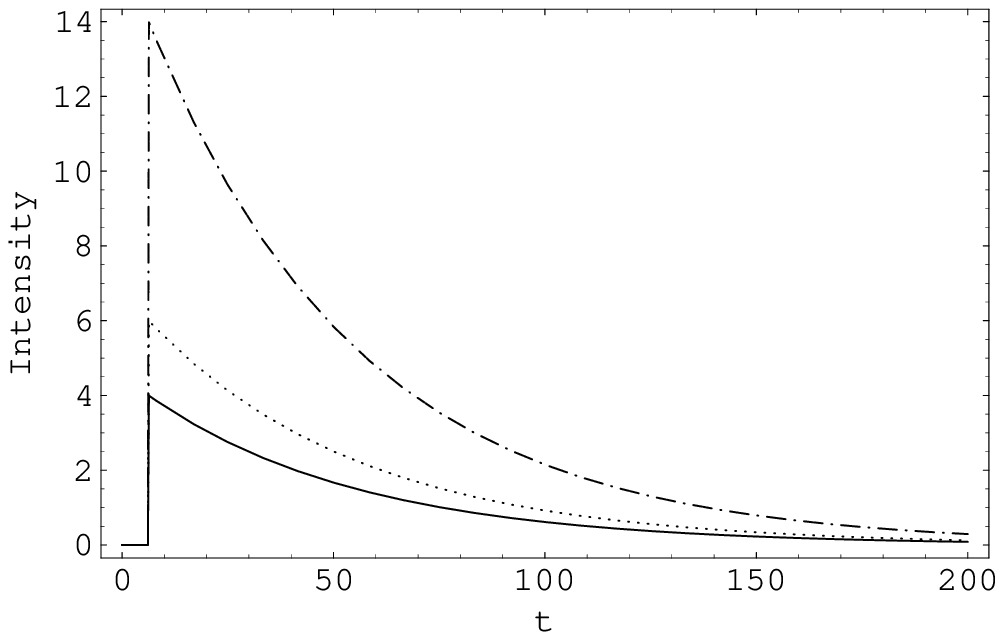}
\end{center}
\caption{Time evolution of light intensity $I(z,t)$ at point $z=2
\pi c/\Omega_1$ for the case that the density matrix of the
initial exciton state is diagonal in Fock representation. The
solid line (Down): $\langle m\rangle _B=0$; The dot line (Middle):
$\langle m\rangle _B=\langle n\rangle _A$; The dashed-dot line
(Up): $\langle m\rangle _B=5\langle n\rangle _A$. The units of
$I(z,t)$ is $\frac{1}{8}\langle n \rangle_A \hbar \Omega_1
\frac{\eta_1}{A}$. $t$ is in units of $1/\Omega_1$. $
\eta_1/2\Omega_1=0.01$, $\Omega_2=0.5\Omega_1$, $\eta_2=\eta_1$.}
\end{figure}

However if $\rho_A$ and $\rho_B$ are non-diagonal in Fock
representation, e.g., the exciton system initially being prepared
in a factorized coherent states $\left|\phi_{ex}\right\rangle
=\left| \alpha \right\rangle _A\otimes\left| \beta \right\rangle
_B$, the light intensity becomes
\end{multicols}
\begin{widetext}
\begin{eqnarray}
I(z,t) &=&\frac{1}{8}\hbar \Omega_{1}\frac{\eta
_{1}}{A}\left\langle n\right\rangle_{A}\Theta(t-z/c)\left\{
4e^{-\eta _{1}(t-z/c)}+4\chi e^{-\eta
_{2}(t-z/c)}\right.  \nonumber\\
&&\left. +8\sqrt{\chi}\cos[(\Omega _{01}-\Omega _{02})(t-z/c)+\phi
]e^{-(\eta _{1}+\eta _{2})(t-z/c)/2}\right\}, \label{eq:intencoh}
\end{eqnarray}
\end{widetext}
\begin{multicols}{2}
\noindent where $\phi$ is the phase difference between the two
coherent states. We find that the last term $\frac 18\hbar \Omega
_3\frac{\eta _3}A\left\{...\right\}$ in Eq. (\ref{eq:intencoh})
gives a nonzero contribution. The detector at $z$ may register an
fluorescence signal oscillating at frequency
$|\Omega_{01}-\Omega_{02}|$. The temporal interference phenomenon
may be observed by using a broadband detector \cite{Meystre}
because with which there is no way to know the photon received by
the detector is emitted from which modes of excitons. This
unknowing for the ``which-way" detection will result in the so
called quantum beat phenomenon.

Form Eq. (\ref{eq:intencoh}), one can easily find that $I(z,t)$ is
oscillating with the time evolution. The detector at $z$ will
receive the first peak after the propagation time $t=z/c$ (In fig.
3, we take $z=2\pi c/\Omega_1$. It is not a necessary choice.).
The influences of $\chi$ on the time evolution of the light
intensity are investigated in Fig. 3, where we have used $\phi=0$,
i.e., zero phase difference between the two coherent states. Our
results show that if one of the two-mode excitons is initially in
vacuum state (say $\beta=0$), we recover the normal exponential
decay of the light intensity. Besides, the amplitudes of the light
intensity become larger with the increase of $\chi$. The result
that the light intensity at initial stage does not go down to zero
(the dot line and the dashed-dot line) is the consequence of
unsymmetrically excitation, i.e., the initially generation of the
two-mode excitons with different mean numbers.
\begin{figure}[hbtp]
\begin{center}
\epsfxsize=7.0cm\epsffile{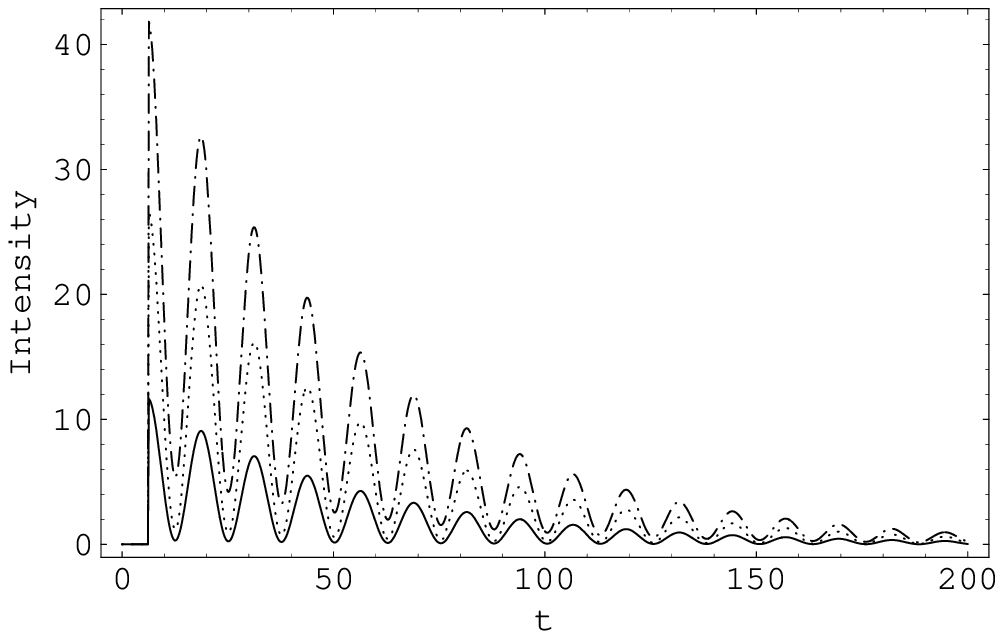}
\end{center}
\caption{Time evolution of light intensity for the case that the
excitons are initially in a factorized coherent state. The Solid
line: $|\beta|^2=|\alpha|^2 $; The Dot line:
$|\beta|^2=5|\alpha|^2 $; The Dashed-dot line: $|\beta|^2=10
|\alpha|^2 $. Other parameters are the same with Figure 2.}
\end{figure}

All our mentioned above are separable state cases, in which the
density matrix of the two-mode exciton system can be factorized
initially. On the other hand, if the system is initially in an
entangled state, e.g., $\left| \phi_{ex} \right\rangle
=\frac{1}{\sqrt{2}}\left(\left| 0 \right\rangle
_A\otimes\left|1\right\rangle _B+\left| 1 \right\rangle
_A\otimes\left|0\right\rangle _B\right)$, one may also observe the
quantum beat phenomenon.

The first-order degree of coherence of the light field
$|g^{(1)}(\tau)|$ as a function of the time delay $\tau$ is
calculated by substituting Eq. (\ref{eq:EzRWA}) and its Hermitian
conjugate into Eq. (19). We find that if one of the two-mode
excitons is initially in vacuum state, e.g., $\left\langle
m\right\rangle_B=0$, then $|g^{(1)}(\tau)|=1$ with regardless of
the another mode state, which implies that single-mode exciton in
the 2D crystal slab emits a complete coherent light.
Mathematically, the complete coherent light is obtained when
$|g^{(1)}(\tau)|=1$. This happens when \cite{Scully}
\begin{equation}
\langle \hat{E}^{(-)}(z,t)\hat{E} ^{(+)}(z,t+\tau)\rangle\propto
{\mathcal E}^*(z,t){\mathcal E}(z,t+\tau),
\end{equation}
where ${\mathcal E}(z,t)=\langle \hat{E} ^{(+)}(z,t)\rangle$,
i.e., two-time correlation function of the total electric field
may be factorized as the ensemble averages of the electric fields.
As a special case, when there is only one mode exciton being
excited initially, one can easily find that the factorized
condition is satisfied. The physical meaning of $g^{(1)}(\tau)$
can be understood by considering the visibility of the
interference fringes, which is proportional to $|g^{(1)}(\tau)|$.
A maximum visibility of the fringes is obtained when
$|g^{(1)}(\tau)|=1$.

If both of the two-mode excitons are populated initially with
$\rho_{ex}=\rho_A \otimes \rho_B$ and both $\rho_A$ and $\rho_B$
being diagonal in Fock representation, the factorized condition of
the two-time correlation function is broken, which leads to
$0<|g^{(1)}(\tau)|<1$. The magnitude of the first-order degree of
coherence for this case is
\end{multicols}
\begin{widetext}
\begin{equation}
|g^{(1)}(z,t;z,t+\tau )|_{z=ct}=\frac{\sqrt{1+2\chi e^{-(\eta
_{2}-\eta _{1})\tau /2}\cos (\Omega _{01}-\Omega _{02})\tau +\chi
^{2}e^{-(\eta _{2}-\eta _{1})\tau }}}{\sqrt{1+\chi +\chi e^{-(\eta
_{2}-\eta _{1})\tau }+\chi ^{2}e^{-(\eta _{2}-\eta _{1})\tau
}}}.\label{eq:g1}
\end{equation}
\end{widetext}
\begin{multicols}{2}
\noindent We find that $|g^{(1)}(\tau)|$ oscillates regularly with
$\tau$, as shown in Fig. 4(a). The magnitude of the first-order
degree of coherence also depends on $\chi$: the larger the degree
of unsymmetrically excitation is, the smaller the oscillating
amplitude of $|g^{(1)}(\tau)|$ is. When $\chi\rightarrow \infty$,
$|g^{(1)}(\tau)|$ tends to $1$. For the case of the excitons being
initially in a factorized coherent state, as shown in Fig. 4(b),
$|g^{(1)}(\tau)|$ is equal to $1$ due to the satisfaction of the
factorized condition.
\begin{figure}[hbtp]
\begin{center}
\epsfxsize=7.0cm\epsffile{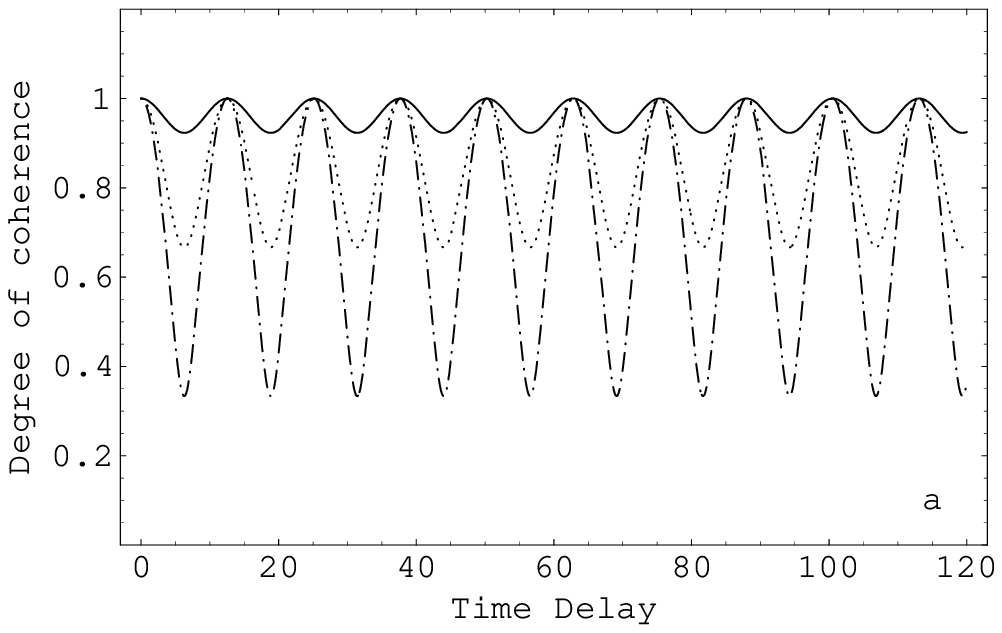}\\[0pt]
\epsfxsize=7.0cm\epsffile{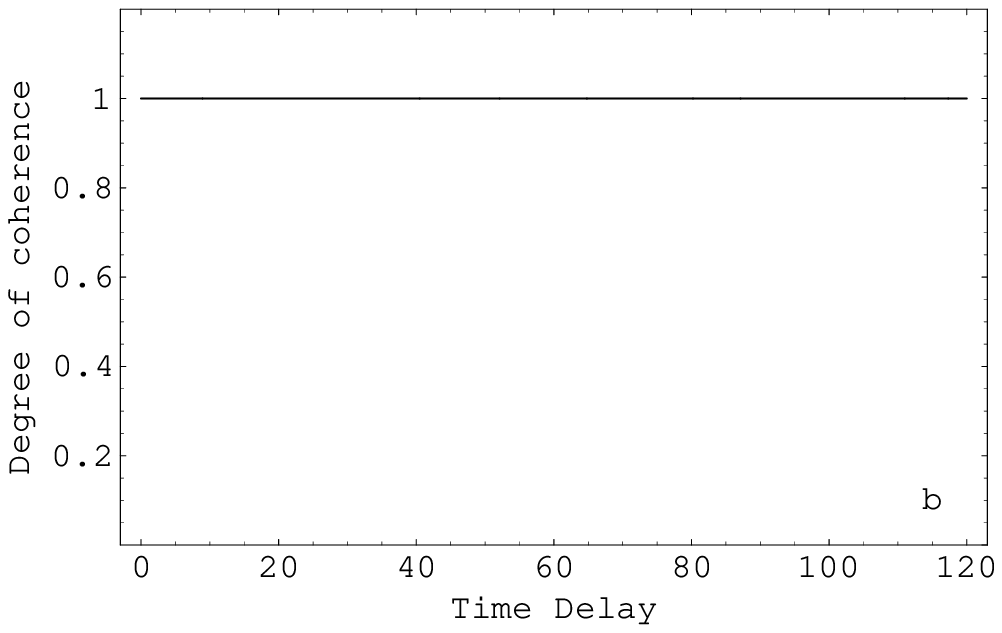}
\end{center}
\caption{$|g^{(1)}(\tau)|$ as a function of time delay $\tau$.
$z=2\pi c/\Omega_1$, $t=2\pi/\Omega_1$, $\eta_1/2\Omega _1=0.01$,
$\Omega_2=0.5\Omega_1$, $\eta _2=\eta_1$. (a) The case that the
density matrix of initial exciton state is diagonal in Fock
representation. The solid line (up) is $\langle m \rangle_B=50
\langle n \rangle_A$; The dot line (middle) is $\langle m
\rangle_B=10\langle n \rangle_A$; and the dashed-dot line (down)
is $\langle m \rangle_B=\langle n \rangle_A$. (b) For a factorized
coherent state case.}
\end{figure}

The second degree of coherence $|g^{(2)}(\tau)|$ as a function of
$\tau$ is also calculated. We find that, when one of the two-mode
excitons is initially in a vacuum state (say $\left\langle
m\right\rangle_B=0$), then: (1) $|g^{(2)}(\tau)|=2$, for the case
of the other mode exciton being in a chaotic state; (2)
$|g^{(2)}(\tau)|=1-1/n$, for the case of the $A_0$-mode excitons
being initially in a Fock state; (3) $|g^{(2)}(\tau)|=1$, for the
coherent state case. If both of the two-mode excitons are
populated initially, we find that $|g^{(2)}(\tau)|$ oscillates
regularly with $\tau$, and the oscillating amplitude of
$|g^{(2)}(\tau)|$ becomes smaller with the increase of $\chi$.

\section{Non-perturbation approach without WWA}
In general, the rotating-wave approximation and Weisskopf-Wigner
approximation are used frequently in quantum optics to study the
fluorescence emitted from a single atom system. However, for the
case of exciton radiance from a multi-atom system, we do not know
exactly if they are valid because of the super-radiance feature of
excitons. In this section, we will restudy the fluorescence of
Frenkel excitons in the V-typed crystal slab by using
non-perturbation approach \cite {Hanamura,Konester1} without the
rotating-wave approximation in the interaction Hamiltonian
$(e/mc)\mathbf{P}\cdot\mathbf{A}$. The self-interaction term $
(e^2/2mc^2){\bf A}^2$ is also included in our model to avoid
un-physical roots in the characteristic equations for the exciton
dispersion relation \cite{Cao}. Both the stimulated emission and
reabsorption effects are taken into account in the new theoretical
treatment. We focus our attention on the radiative decay rates of
the excitons and the light intensities for various exciton initial
states to compare with our previous results in Sec. IV.

The interaction Hamiltonian $H_1$ between the Frenkel excitons and
photons without RWA has been given in Eq. (8) in section II. Our
further discussion will also take the self-interaction term
$\frac{e^2}{2mc^2}\mathbf{A}^2$ into account. In second
quantization, this self-interaction Hamiltonian is written as
\begin{equation}
\hat{H}_2=\hbar \sum_{q,q^{\prime };l}f(q,q^{\prime })\left[
\hat{a}_q+\hat{a}_{-q}^{\dagger } \right] \left[\hat{a}_{q^{\prime
}}+\hat{a}_{-q^{\prime }}^{\dagger }\right] e^{i(q+q^{\prime
})la},
\end{equation}
in which
\begin{equation}
f(q,q^{\prime })=\frac{N_L\pi e^2}{mcV\sqrt{|qq^{\prime }|}}\sum_{\lambda
,\lambda ^{\prime }}\mathbf{e}_{q\lambda }\cdot \mathbf{e}_{q^{\prime
}\lambda ^{\prime }},
\end{equation}
where $\mathbf{e}_{q\lambda }$ is the unit polarization vector of
the $(q,\lambda)$-mode photon with $\lambda =1,2$. The sum
$\sum_{\lambda ,\lambda ^{\prime }}\mathbf{e}_{q\lambda }\cdot
\mathbf{e}_{q^{\prime }\lambda ^{\prime }}$ had been given for the
two-level atom case \cite {Milonni,Cao} with certain assumptions.
Following them we can give a similar result for the V-type
three-level case (the detailed calculations are presented in
Appendix B):
\begin{eqnarray}
f(q,q^{\prime })&=&\frac{2\pi N_L}{c\hbar V\sqrt{|qq^{\prime }|}}\left[
\Omega _1d_1^2+\Omega _2d_2^2\right]  \nonumber \\
&=&\frac 1N\left[ \frac 1{\Omega _1}G_1(q)G_1(q^{\prime })+\frac 1{\Omega
_2} G_2(q)G_2(q^{\prime })\right].
\end{eqnarray}
By making use of
\begin{equation}
\sum_le^{i(q+q^{\prime })la}=N\sum_kO(q^{\prime }-k)O(k+q),
\end{equation}
we can get the self-interaction Hamiltonian of photon $\hat{H}_2$.
The Heisenberg equations for exciton and photon operators can be
obtained from the total interaction Hamiltonian
$\hat{H}_{int}=\hat{H}_1+\hat{H}_2$ (Eq. (C1) in Appendix C).
Performing half-side Fourier transformation (HSFT)
\begin{equation}
A_q(\omega) =\int_0^\infty \hat{A}_q(t)e^{i\omega t}dt
\end{equation}
on both sides of the equations, four algebraic equations about
$A_{k}(\omega )$, $B_{k}(\omega )$, $a_{q}(\omega)$, and
$a_{-q}^{\dagger}(\omega)$ are obtained as listed in Appendix C.
Combining these equations we get
\begin{eqnarray}
&&a_q(\omega )-a_{-q}^{\dagger }(\omega) \nonumber \\
&=&\frac{2\omega }{\Omega _1 (\omega ^2-q^2c^2)}G_1(q)\sum_kO(k-q)
\nonumber\\
&&\times\left[\omega \left[A_k(\omega)-A_{-k}^{\dagger}(\omega
)\right]-i\left[A_k(0)-A_{-k}^{\dagger}(0)\right]\right]  \nonumber \\
&&+\frac{2\omega }{\Omega_2 (\omega ^2-q^2c^2)}G_2(q)\sum_kO(k-q)
\nonumber\\
&&\times \left[\omega \left[B_k(\omega )-B_{-k}^{\dagger }(\omega
)\right]
-i\left[B_k(0)-B_{-k}^{\dagger}(0)\right] \right]  \nonumber \\
&&+\frac{i}{\omega -|q|c}a_q(0)-\frac{i}{\omega
+|q|c}a_{-q}^{\dagger}(0),\label{eq:aqaq}
\end{eqnarray}
for the electric field of the fluorescence. In order to calculate
it, one need to know $A_k(\omega )-A_{-k}^{\dagger }(\omega )$ and
$B_k(\omega )-B_{-k}^{\dagger }(\omega )$, which are determined by
$a_q(\omega )+a_{-q}^{\dagger }(\omega)$. In appendix C, two
coupled equations for $A_k(\omega )-A_{-k}^{\dagger }(\omega )$
and $B_k(\omega )-B_{-k}^{\dagger }(\omega )$ are presented in
detail. We solve these two equations consistently and replace them
into Eq. (\ref{eq:aqaq}). Then we get the electric field by the
inverse HSFT
\begin{equation}
E(z,t)=\frac 1{2\pi }\int_{-\infty +i\epsilon }^{\infty +i\epsilon
}E(z,\omega )e^{-i\omega t}d\omega ,\label{eq:Ez}
\end{equation}
where
\begin{eqnarray}
E(z,\omega ) &=&i\sum_q\sqrt{\frac{2\pi |q|c\hbar }V}\left[ a_q(\omega
)-a_{-q}^{\dagger }(\omega )\right] e^{iqz}  \nonumber \\
&=&i\int_{-\infty }^\infty \sqrt{\frac{|q|c\hbar L}{2\pi A}}\left[
a_q(\omega )-a_{-q}^{\dagger }(\omega )\right]
e^{iqz}dq,\label{eq:Ew}
\end{eqnarray}
for sufficiently large $L$.

From the explicit form of the electric field operator, the
radiative decay rates of Frenkel excitons can be solved. For the
sake of simplicity, we will restrict ourselves to the monolayer
case $N=1$, in which only zero wave vector of excitons are
involved in our model. Solving the two coupled equations for
$A_0(\omega )-A_0^{\dagger }(\omega )$ and $B_0(\omega
)-B_0^{\dagger }(\omega )$ (see Eqs. (\ref{eq:CoupEq1}) and
(\ref{eq:CoupEq2}) in Appendix C), we get two independent
equations
\end{multicols}
\begin{widetext}
\begin{eqnarray}
&&\left\{ \left[ \omega ^{2}-\Omega _{1}^{2}-\frac{2\omega
^{2}}{\Omega _{1}} F_{00}^{(1)}(\omega )\right] \left[ \omega
^{2}-\Omega _{2}^{2}-\frac{
2\omega ^{2}}{\Omega _{2}}F_{00}^{(2)}(\omega )\right] \right. \nonumber \\
&&\left. -\frac{2\omega ^{2}}{\Omega _{1}}F_{00}^{(3)}(\omega
)\frac{2\omega ^{2}}{\Omega _{2}}F_{00}^{(3)}(\omega )\right\}
\left[ A_{0}(\omega
)-A_{0}^{\dagger }(\omega )\right]   \nonumber \\
&=&i\left[ \omega ^{2}-\Omega _{2}^{2}-\frac{2\omega ^{2}}{\Omega
_{2}} F_{00}^{(2)}(\omega )\right] \left[ (\omega +\Omega
_{1})A_{0}(0)-(\omega
-\Omega _{1})A_{0}^{\dagger }(0)\right]   \nonumber \\
&&+2i\omega F_{00}^{(3)}(\omega )\left[ (\omega +\Omega
_{2})B_{0}(0)-(\omega -\Omega _{2})B_{0}^{\dagger }(0)\right] \nonumber \\
&&-i\frac{2\omega }{\Omega _{1}}F_{00}^{(1)}(\omega )\left( \omega
^{2}-\Omega _{2}^{2}\right) \left[ A_{0}(0)-A_{0}^{\dagger
}(0)\right] ,\label{eq:CloseEq1}
\end{eqnarray}
and
\begin{eqnarray}
&&\left\{ \left[ \omega ^{2}-\Omega _{2}^{2}-\frac{2\omega
^{2}}{\Omega _{2}} F_{00}^{(2)}(\omega )\right] \left[ \omega
^{2}-\Omega _{1}^{2}-\frac{
2\omega ^{2}}{\Omega _{1}}F_{00}^{(1)}(\omega )\right] \right.\nonumber \\
&&\left. -\frac{2\omega ^{2}}{\Omega _{2}}F_{00}^{(3)}(\omega
)\frac{2\omega ^{2}}{\Omega _{1}}F_{00}^{(3)}(\omega )\right\}
\left[ B_{0}(\omega
)-B_{0}^{\dagger }(\omega )\right]   \nonumber \\
&=&i\left[ \omega ^{2}-\Omega _{1}^{2}-\frac{2\omega ^{2}}{\Omega
_{1}} F_{00}^{(1)}(\omega )\right] \left[ (\omega +\Omega
_{2})B_{0}(0)-(\omega
-\Omega _{2})B_{0}^{\dagger }(0)\right]   \nonumber \\
&&+2i\omega F_{00}^{(3)}(\omega )\left[ (\omega +\Omega
_{1})A_{0}(0)-(\omega -\Omega _{1})A_{0}^{\dagger }(0)\right]\nonumber \\
&&-i\frac{2\omega }{\Omega _{2}}F_{00}^{(2)}(\omega )(\omega
^{2}-\Omega _{1}^{2})\left[ B_{0}(0)-B_{0}^{\dagger }(0)\right]
,\label{eq:CloseEq2}
\end{eqnarray}
\end{widetext}
\begin{multicols}{2}
\noindent where
\begin{equation}
F_{00}^{(j)}(\omega )=-i\frac{\Omega_j\eta _j}{2\omega }\text{,
}\eta _j=\frac{af_j^2}{2c}\text{, }f_j^2=\frac{8\pi \Omega
_jd_j^2}{\hbar a^3},
\end{equation}
for $j=1\text{, }2\text{, }3$. $F^{(j)}_{00}(\omega)$ for
$j=1\text{, }2$ give the exciton wave function's overlap for the
$A_0$- and $B_0$-mode exciton, respectively, however
$F^{(3)}_{00}(\omega)$ is that between the two modes. Note that
MPP has been taken into account in Eqs. (\ref{eq:CloseEq1}) and
(\ref{eq:CloseEq2}). Then, we get the same characteristic
equations for decay rates and frequency shifts of the two exciton
modes $A_0$ and $B_0$
\begin{equation}
\zeta (\omega )=0, \label{eq:CharaEq}
\end{equation}
where
\begin{eqnarray}
\zeta (\omega ) &=&(\omega ^2-\Omega _1^2+i\eta _1\omega )(\omega
^2-\Omega _2^2+i\eta _2\omega )+\eta _1\eta _2\omega
^2.\nonumber\\\label{eq:zeta}
\end{eqnarray}
The roots of Eq. (\ref{eq:CharaEq}) can be solved exactly,
\begin{eqnarray}
\omega _1 &=&\Omega _{01}-i\Gamma _1/2,\omega _2=-\Omega
_{01}-i\Gamma _1/2,\nonumber \\
\omega _3 &=&\Omega _{02}-i\Gamma _2/2, \omega _4=-\Omega
_{02}-i\Gamma _2/2,\label{eq:roots}
\end{eqnarray}
which determine the poles of $A_0(\omega )-A_0^{\dagger }(\omega
)$ and $ B_0(\omega )-B_0^{\dagger }(\omega )$. The imaginary
parts of the roots are the decay rates of the excitons, and the
real parts are the renormalized physical frequencies. For the case
$\Omega _1=\Omega _2=\Omega $, i.e., the degenerate case, we get
\begin{eqnarray}
\omega_1 &=&\Omega ,\omega_3=\Omega_0-i\left(\eta_1+\eta _2\right)
/2,\nonumber \\
\omega_2&=&-\Omega,\omega_4=-\Omega_0-i\left(\eta_1+\eta_2\right)/2,
\end{eqnarray}
where
\begin{equation}
\Omega_0=\sqrt{\Omega ^2-\frac 14\left( \eta _1+\eta _2\right) ^2}
\approx \Omega \left[ 1-\frac{\left( \eta _1+\eta _2\right)
^2}{8\Omega ^2}\right].
\end{equation}
For the non-degenerate case $\Omega _1\neq \Omega _2$, we assume
that the physical roots of the characteristic equation are not far
away from $\pm \Omega _j$ due to the condition $\eta_j\ll\Omega
_j$. Therefore, one can expand $\omega$ up to third-order of $\eta
_j/\Omega_j$, and get
\begin{eqnarray}
\Omega _{01} &=&\Omega _1\left[ 1-\frac{\eta _1^2}{8\Omega
_1^2}-\frac 12
\frac{\eta _1\eta _2}{\Omega _1^2-\Omega _2^2}\right] ,  \nonumber \\
\Omega _{02} &=&\Omega _2\left[ 1-\frac{\eta _2^2}{8\Omega
_2^2}-\frac 12
\frac{\eta _1\eta _2}{\Omega _2^2-\Omega _1^2}\right] ,  \nonumber \\
\Gamma _1 &=&\eta _1\left[ 1-\frac{\eta _2\Omega _1^2-\eta
_1\Omega _2^2}{
\left( \Omega _1^2-\Omega _2^2\right) ^2}\eta _2\right] ,  \nonumber \\
\Gamma _2 &=&\eta _2\left[ 1-\frac{\eta _1\Omega _2^2-\eta
_2\Omega _1^2}{ \left( \Omega _2^2-\Omega _1^2\right) ^2}\eta
_1\right],\label{eq:rate}
\end{eqnarray}
where $\eta _j$ with $j=1,2$ are given in Eqs. (\ref{eq:eta1}) and
(\ref{eq:eta2}).

We compare our results with the two-level lattice atoms case
\cite{Cao}, in which the radiative decay rate of excitons is
$\eta=3\left(\frac{\pi \lambda^2}{a^2} \right)\gamma$, where
$\gamma$ is the decay rate of an isolated lattice atom,
$\lambda=c/\Omega$ denotes the reduced wavelength, and physical
frequency is $\Omega_{0}=\Omega \left[ 1-\frac{\eta ^2}{8\Omega
^2}\right]$. However, we see from Eq. (\ref{eq:rate}), for the
three-level lattice atom case, the decay rates $\Gamma _j$ are
different from the two-level atom case $\eta _j$ by a third-order
correction, which comes from MPP (including stimulation emission,
absorption and other high-order processes). Here $\Omega
_{01}-\Omega _1$ and $\Omega _{02}-\Omega _2$ denote the frequency
shifts for the $A_0$-mode and $B_0$-mode exciton, respectively.
For the two-level case the frequency shift is $\frac{\eta ^2}{
8\Omega ^2}$. Whereas, for three-level case considered here, the
phase shifts for the two-mode exciton are $\frac{\eta
_i^2}{8\Omega _i^2}+\frac 12 \frac{\eta _i\eta _j}{\Omega
_i^2-\Omega _j^2}$, $i\text{, }j=1\text{, }2$. The last term in
the frequency shifts is a second-order correction, which can be
adjusted by tuning the energy spacing of the upper two levels of
the lattice atoms.

In the remainder of this paper, we will present the explicit
expressions of the electric field operator. The time evolution and
spatial distribution of the light intensity are calculated to
compare with the results of RWA in Sec. IV.

From Eqs. (\ref{eq:CloseEq1}) and (\ref{eq:CloseEq2}), the
explicit expressions for $A_0(\omega )-A_0^{\dagger }(\omega )$
(Eq. (\ref{eq:sol1})) and $B_0(\omega )-B_0^{\dagger }(\omega )$
(Eq. (\ref{eq:sol2})) are obtained. Replacing them into Eq.
(\ref{eq:aqaq}), we get the explicit light field operator in terms
of the initial exciton field operators
\begin{eqnarray}
&&a_q(\omega )-a_{-q}^{\dagger }(\omega )  \nonumber \\
&=&\frac{2i\omega G_1(q)(\omega ^2-\Omega _2^2)}{(\omega ^2-q^2c^2)\zeta
(\omega )}\left[ (\omega +\Omega _1)A_0(0)+(\omega -\Omega _1)A_0^{\dagger
}(0)\right]  \nonumber \\
&+&\frac{2i\omega G_2(q)(\omega ^2-\Omega _1^2)}{(\omega
^2-q^2c^2)\zeta (\omega )}\left[ (\omega +\Omega _2)B_0(0)+(\omega
-\Omega _2)B_0^{\dagger }(0)\right],\nonumber\\
\end{eqnarray}
where we have neglected the quantum noise terms proportional to
$a_q(0)$ and $a_{-q}^{\dagger }(0)$ in the above calculation.
Substituting $a_q(\omega )-a_{-q}^{\dagger }(\omega )$ into Eq.
(\ref{eq:Ew}) and choosing a proper contour integration in the
upper complex $q$ plane, we get $E(z,\omega )$ in the positive $z$
region outside the crystal slab as the following,
\begin{eqnarray}
&&E(z,\omega )  \nonumber \\
&=&i\sqrt{\frac{\pi \hbar \Omega _1\eta _1}{cA}}\frac{\omega
^2-\Omega _2^2}{ \zeta (\omega )}\left[ (\omega +\Omega
_1)A_0(0)\right.\nonumber\\
&&\left.+(\omega -\Omega_1)A_0^{\dagger }(0)\right] e^{i\frac \omega cz}
\nonumber \\
&+&i\sqrt{\frac{\pi \hbar \Omega _2\eta _2}{cA}}\frac{\omega
^2-\Omega _1^2} {\zeta (\omega )}\left[ (\omega +\Omega
_2)B_0(0)\right.\nonumber\\
&&\left.+(\omega -\Omega _2)B_0^{\dagger }(0)\right]e^{i\frac
\omega cz},\nonumber\\
\end{eqnarray}
where $A$ is the area of the layer. As mentioned above, we choose
the normalization volume for the photon as $AL$.

The electric field $E(z,t)$ in the positive $z$ region outside the
crystal slab can be calculated by Eq. (\ref{eq:Ez}). For the case
$z-ct>0$ ($<0$), the contour integral over $d\omega$ can be
performed by choosing the integration path in the upper (lower)
complex $\omega$ plane. The result is that $E(z,t)=0$, for
$z-ct>0$; but for $z-ct<0$,
\begin{equation}
E(z,t)={\mathcal E^{(+)}}(z,t)+\mathrm{H.c.}.
\end{equation}
The explicit expression for ${\mathcal E^{(+)}}(z,t)$ in Appendix
C shows that, in positive $z$ region, the electric field generated
by the two-mode exciton is damped exponentially with two
eigen-decay rates $\Gamma _1$ and $\Gamma _2$. The corresponding
eigen-modes are linear combinations of $A_0$- and $B_0$-mode. The
electric field in negative $z$ region can also be derived
similarly and shows that it is a damped wave propagating in the
negative $z$ direction. Since all the roots of the characteristic
equation are in lower half plane of complex $\omega$ plane, then
$E(z,t)=0$, for $z-ct>0$, so our result is reasonable and obeys
the causal rule.

Similarly, we can also calculate the positive frequency part of
the electric field operator, which is determined by $a_q(\omega)$.
The explicit expression of $a_q(\omega)$ can be obtained by
substituting Eqs. (\ref{eq:sol1}) and (\ref{eq:sol2}) into Eq.
(\ref{eq:aq}) (see details in Appendix C.) and omitting quantum
noise terms. We get
\begin{eqnarray}
&&(\omega -|q|c)a_q(\omega )\nonumber \\
&=&iG_1(q)\frac {\omega^2-\Omega_2^2} {\zeta (\omega )}
\left[(\omega+\Omega_1)A_0(0)+(\omega-\Omega_1)A_0^{\dagger
}(0)\right] \nonumber \\
&+&iG_2(q)\frac {\omega^2-\Omega_1^2} {\zeta (\omega )}\left[
(\omega+\Omega_2)B_0(0)+(\omega-\Omega_2)B_0^{\dagger }(0)\right].
\nonumber\\
\end{eqnarray}
Therefore, we get
\begin{eqnarray}
E^{(+)}(z,t) &=&\frac i{2\pi }\int_{-\infty }^\infty
dq\int_{-\infty +i\epsilon }^{\infty +i\epsilon }d\omega
\sqrt{\frac{|q|c\hbar
L}{2\pi A}} a_q(\omega )e^{i(qz-\omega t)}\nonumber\\
&=&\frac{1}{2\pi }\int_{-\infty }^{\infty }dqe^{iqz}\int_{-\infty
+i\epsilon }^{\infty +i\epsilon }d\omega
\frac{e^{-i\omega t}}{\omega _{q}-\omega } \nonumber\\
&&\times \left\{ \sqrt{\frac{c\hbar \Omega _{1}\eta _{1}}{4\pi
A}}\frac{ (\omega ^{2}-\Omega _{2}^{2})}{\zeta (\omega )}\left[
(\omega +\Omega_{1})\right. \right.  \nonumber\\
&&\times \left. A_{0}(0)+(\omega -\Omega _{1})A_{0}^{\dagger
}(0)\right]\nonumber\\
&&+\left. \sqrt{\frac{c\hbar \Omega _{2}\eta _{2}}{4\pi
A}}\frac{(\omega ^{2}-\Omega _{1}^{2})}{\zeta (\omega )}\left[
(\omega +\Omega _{2})\right.
\right.  \nonumber\\
&&\times \left. \left. B_{0}(0)+(\omega -\Omega
_{2})B_{0}^{\dagger }(0) \right] \right\}.
\end{eqnarray}

A straightforward calculations give the following results:
\begin{equation}
E^{(+)}(z,t)=0\text{, for $t<0$,}
\end{equation}
\begin{eqnarray}
E^{(+)}(z,t)&=&\sqrt{\frac{\pi \hbar \Omega _1\eta _1}{4Ac}}\left[
F_A^{(+)}(z,t)A_0(0)\right.\nonumber\\
&&\left.+F_A^{(-)}(z,t)A_0^{\dagger }(0)\right]  \nonumber \\
&+&\sqrt{\frac{\pi \hbar \Omega _2\eta _2}{4Ac}}\left[
F_B^{(+)}(z,t)B_0(0)\right.\nonumber\\
&&\left.+F_B^{(-)}(z,t)B_0^{\dagger }(0)\right]\text{, for $t>0$,}
\end{eqnarray}
in which the time-dependent coefficients $F_A^{(\pm )}(z,t)$ and
$F_B^{(\pm )}(z,t)$ are
\end{multicols}
\begin{widetext}
\begin{eqnarray}
F_A^{(\pm )}(z,t) &=&\frac i\pi \int_0^{\infty}d\omega _q
2\left[\frac{\left( \omega _q^2-\Omega _2^2\right) \left( \omega
_q\pm \Omega _1\right) e^{-i\omega _qt}}{\left( \omega _q-\omega
_1\right) \left( \omega _q-\omega _2\right) \left( \omega
_q-\omega _3\right) \left( \omega _q-\omega
_4\right) } \right. \nonumber\\
&&\left.-\frac{\left( \omega _1^2-\Omega _2^2\right) \left( \omega
_1\pm \Omega _1\right) e^{-i\omega _1t}}{\left( \omega _q-\omega
_1\right) \left( \omega _1-\omega _2\right) \left( \omega
_1-\omega _3\right) \left( \omega
_1-\omega _4\right) } \right. \nonumber\\
&&\left.-\frac{\left( \omega _2^2-\Omega _2^2\right) \left( \omega
_2\pm \Omega _1\right) e^{-i\omega _2t}}{\left( \omega _q-\omega
_2\right) \left( \omega _2-\omega _1\right) \left( \omega
_2-\omega _3\right) \left( \omega
_2-\omega _4\right) } \right.\nonumber\\
&&\left.-\frac{\left( \omega _3^2-\Omega _2^2\right) \left( \omega
_3\pm \Omega _1\right) e^{-i\omega _3t}}{\left( \omega _q-\omega
_3\right) \left( \omega _3-\omega _1\right) \left( \omega
_3-\omega _2\right) \left( \omega
_3-\omega _4\right) }\right.\nonumber\\
&&\left.-\frac{\left( \omega _4^2-\Omega _2^2\right) \left( \omega
_4\pm \Omega _1\right) e^{-i\omega _4t}}{\left( \omega _q-\omega
_4\right) \left( \omega _4-\omega _1\right) \left( \omega
_4-\omega _2\right) \left( \omega _4-\omega _3\right) }\right]\cos
\left( \omega _q\frac zc\right) ,\label{eq:fa}
\end{eqnarray}
\begin{eqnarray}
F_B^{(\pm )}(z,t) &=&\frac i\pi \int_0^{\infty}d\omega _q 2
\left[\frac{\left( \omega _q^2-\Omega _1^2\right) \left( \omega
_q\pm \Omega _2\right) e^{-i\omega _qt}}{\left( \omega _q-\omega
_1\right) \left( \omega _q-\omega _2\right) \left( \omega
_q-\omega _3\right) \left( \omega _q-\omega
_4\right) }  \right.\nonumber \\
&&\left.-\frac{\left( \omega _1^2-\Omega _1^2\right) \left( \omega
_1\pm \Omega _2\right) e^{-i\omega _1t}}{\left( \omega _q-\omega
_1\right) \left( \omega _1-\omega _2\right) \left( \omega
_1-\omega _3\right) \left( \omega
_1-\omega _4\right) }\right.  \nonumber \\
&&\left.-\frac{\left( \omega _2^2-\Omega _1^2\right) \left( \omega
_2\pm \Omega _2\right) e^{-i\omega _2t}}{\left( \omega _q-\omega
_2\right) \left( \omega _2-\omega _1\right) \left( \omega
_2-\omega _3\right) \left( \omega
_2-\omega _4\right) } \right. \nonumber \\
&&\left.-\frac{\left( \omega _3^2-\Omega _1^2\right) \left( \omega
_3\pm \Omega _2\right) e^{-i\omega _3t}}{\left( \omega _q-\omega
_3\right) \left( \omega _3-\omega _1\right) \left( \omega
_3-\omega _2\right) \left( \omega
_3-\omega _4\right) } \right. \nonumber \\
&&\left.-\frac{\left( \omega _4^2-\Omega _1^2\right) \left( \omega
_4\pm \Omega _2\right) e^{-i\omega _4t}}{\left( \omega _q-\omega
_4\right) \left( \omega _4-\omega _1\right) \left( \omega
_4-\omega _2\right) \left( \omega _4-\omega _3\right) }\right]\cos
\left( \omega _q\frac zc\right), \label{eq:fb}
\end{eqnarray}
\end{widetext}
\begin{multicols}{2}
\noindent where $\omega _j$, for $j=1,2,3,4$, are four roots of
the characteristic equation of Eq. (\ref{eq:CharaEq}). We again
need to carry out the integration over $d\omega_q$ in the above
two equations. It is easy to prove that $F_{A}^{(+)}(z,t)$ can be
simplified as
\begin{eqnarray}
F_{A}^{(+)}(z,t) &=&\frac{i}{\pi }\int_{0}^{\infty }d\omega
_{q}2\cos (\omega _{q}\frac{z}{c})\left[ c_{1}\frac{e^{-i\omega
_{q}t}-e^{-i\omega
_{1}t}}{\omega _{q}-\omega _{1}}\right.\nonumber \\
&&+\left. c_{2}\frac{e^{-i\omega _{q}t}-e^{-i\omega _{2}t}}{\omega
_{q}-\omega _{2}}+c_{3}\frac{e^{-i\omega _{q}t}-e^{-i\omega
_{3}t}}{\omega_{q}-\omega _{3}}\right.\nonumber \\
&&+\left. c_{4}\frac{e^{-i\omega _{q}t}-e^{-i\omega _{4}t}}{\omega
_{q}-\omega _{4}}\right] ,
\end{eqnarray}
where the four coefficients are
\begin{eqnarray}
c_{1} &=&\frac{\left( \omega _{1}^{2}-\Omega _{2}^{2}\right)
\left( \omega _{1}+\Omega _{1}\right) }{\left( \omega _{1}-\omega
_{2}\right) \left( \omega _{1}-\omega _{3}\right) \left( \omega
_{1}-\omega _{4}\right) }, \nonumber \\
c_{2} &=&\frac{\left( \omega _{2}^{2}-\Omega _{2}^{2}\right)
\left( \omega _{2}+\Omega _{1}\right) }{\left( \omega _{2}-\omega
_{1}\right) \left( \omega _{2}-\omega _{3}\right) \left( \omega
_{2}-\omega _{4}\right) },\nonumber \\
c_{3} &=&\frac{\left( \omega _{3}^{2}-\Omega _{2}^{2}\right)
\left( \omega _{3}+\Omega _{1}\right) }{\left( \omega _{3}-\omega
_{1}\right) \left( \omega _{3}-\omega _{2}\right) \left( \omega
_{3}-\omega _{4}\right) },\nonumber \\
c_{4} &=&\frac{\left( \omega _{4}^{2}-\Omega _{2}^{2}\right)
\left( \omega _{4}+\Omega _{1}\right) }{\left( \omega _{4}-\omega
_{1}\right) \left( \omega _{4}-\omega _{2}\right) \left( \omega
_{4}-\omega _{3}\right) }.
\end{eqnarray}
The above integral over $d\omega _{q}$ can be done as Sec. IV. The
explicit expressions of $F_{A}^{(-)}(z,t)$ and $F_{B}^{(\pm
)}(z,t)$ can also be obtained with the same calculations. Our
results also confirm the causal rule.

For an arbitrary exciton initial state $\left| \phi_{ex}
\right\rangle$, the light intensity is defined by Eq. (18). When
the excitons are initially in a state with density matrix
$\rho(0)=\rho_A \otimes \rho_B$, and both $\rho_A$ and $\rho_B$
diagonal in Fock representation, the light intensity is
\begin{eqnarray}
I(z,t) &=&\frac 18\hbar \Omega _1\frac{\eta _1}A\left\langle
n\right\rangle _A\left[ \left| F_A^{(+)}(z,t)\right| ^2+\left|
F_A^{(-)}(z,t)\right| ^2
\right]  \nonumber \\
&+&\frac 18\hbar \Omega _2\frac{\eta _2}A\left\langle
m\right\rangle _B\left[ \left| F_B^{(+)}(z,t)\right| ^2+\left|
F_B^{(-)}(z,t)\right| ^2\right],\nonumber\\
\end{eqnarray}
which is shown in Fig. 5. When $\langle m\rangle _B=0$, the solid
line, our result will go back to the two-level lattice atoms case
\cite{Cao}. From the dot line ($\left\langle
m\right\rangle_B/\left\langle n\right\rangle_A=1$) and dashed-dot
line ($\left\langle m\right\rangle_B/\left\langle
n\right\rangle_A=5$) of Fig. 5, we find that with the increase of
$\chi$, the amplitude of the light intensity becomes higher, which
is the same with that of obtained by using RWA (Fig. 2). Contrary
to the results of Sec. IV, however, when we consider it without
RWA, the light intensity does not decay exponentially but in an
irregular way due to the existence of counter-rotating terms in
$E^{(+)}(z,t)$. Besides, it's also deserved to mentioned that the
contributions of the non-rotating terms do not appear as quivers
presented in Ref. \cite{Cao}.
\vskip 0.1cm
\begin{figure}[hbtp]
\begin{center}
\epsfxsize=7.0cm\epsffile{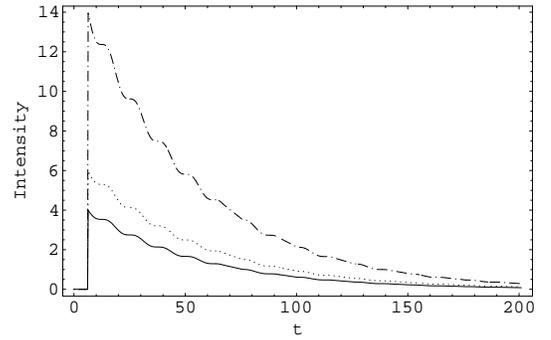}
\end{center}
\caption{Time evolution of light intensity $I(z,t)$ at point $z=2
\pi c/\Omega_1$ for the case that the density matrix of the
initial exciton state is diagonal in Fock representation. The
solid line (Down): $\langle m\rangle _B=0$; The dot line (Middle):
$\langle m\rangle _B=\langle n\rangle _A$; The dashed-dot line
(Up): $\langle m\rangle _B=5\langle n\rangle _A$. Other parameters
are the same with Fig. 2.}
\end{figure}

For the case that the two-mode excitons are initially in a
factorized coherent states, the light intensity becomes
\end{multicols}
\begin{widetext}
\begin{eqnarray}
I(z,t) &=&\frac 18\hbar \Omega _1\frac{\eta _1}A\left[\left| \alpha \right|
^2\left| F_A^{(+)}(z,t)\right| ^2+\left| \alpha \right| ^2\left|
F_A^{(-)}(z,t)\right| ^2\right.  \nonumber \\
&&\left.+(\alpha ^{*})^2F_A^{(+)*}(z,t)F_A^{(-)}(z,t)+\alpha
^2F_A^{(-)*}(z,t)F_A^{(+)}(z,t)\right]  \nonumber \\
&&+\frac 18\hbar \Omega _2\frac{\eta _2}A\left[\left| \beta \right| ^2\left|
F_B^{(+)}(z,t)\right| ^2+\left| \beta \right| ^2\left| F_B^{(-)}(z,t)\right|
^2 \right.  \nonumber \\
&&\left.+\left( \beta ^{*}\right) ^2F_B^{(+)*}(z,t)F_B^{(-)}(z,t)+\beta
^2F_B^{(-)*}(z,t)F_B^{(+)}(z,t)\right]  \nonumber \\
&&+\frac 18\hbar \Omega _3\frac{\eta _3}A\left[\alpha ^{*}\beta
F_A^{(+)*}(z,t)F_B^{(+)}(z,t)+\alpha ^{*}\beta
^{*}F_A^{(+)*}(z,t)F_B^{(-)}(z,t) \right.  \nonumber \\
&&\left.+\alpha \beta F_A^{(-)*}(z,t)F_B^{(+)}(z,t)+\alpha \beta
^{*}F_A^{(-)*}(z,t)F_B^{(-)}(z,t)+C.c.\right],
\label{eq:intencohwithoutrwa}
\end{eqnarray}
\end{widetext}
\begin{multicols}{2}
\noindent where the last term $\frac 18\hbar \Omega _3\frac{\eta
_3}A[...]$ gives a temporal interference term. The effects of
phase difference between the two coherent states on the light
intensity are studied in Fig. 6. We find that the first peak (at
$t=z/c$) of the light intensity becomes lower in amplitude with
the increase of the phase difference from $\phi=0$ to $\phi=\pi$
(monotonic regime). In fact, the whole curve will be shifted left
with the increase of the phase difference within the monotonic
regime, which leads to the magnitude of the peak (at $t=z/c$)
becomes lower (comparing Fig. 6(b) and Fig. 6(c) with Fig. 6(a)).
When $\phi=\pi$, $I(z,t=z/c)$ tends to zero, i.e., the first peak
is disappeared (see Fig. 6(c)). Our results also show that both
the phase difference and the degree of unsymmetrically excitation
do not effect the oscillation frequency of the light intensity,
which is determined by the exciton splitting. The oscillation
behavior in the light intensity may take place as long as
$|\Omega_{01}-\Omega_{02}|>\Gamma_{j}$, i.e., the exciton
splitting is greater than the natural linewidth of exiton, so that
one can observe the beating phenomenon within the lifetime of the
exciton.
\begin{figure}[hbtp]
\begin{center}
\epsfxsize=7.0cm\epsffile{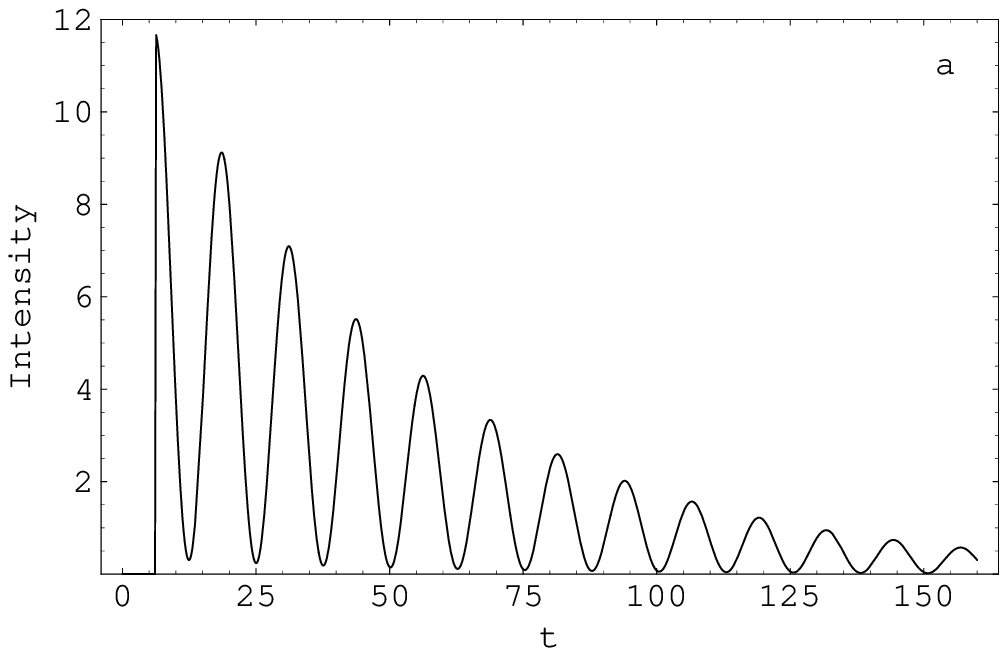}\\[0pt]
\epsfxsize=7.0cm\epsffile{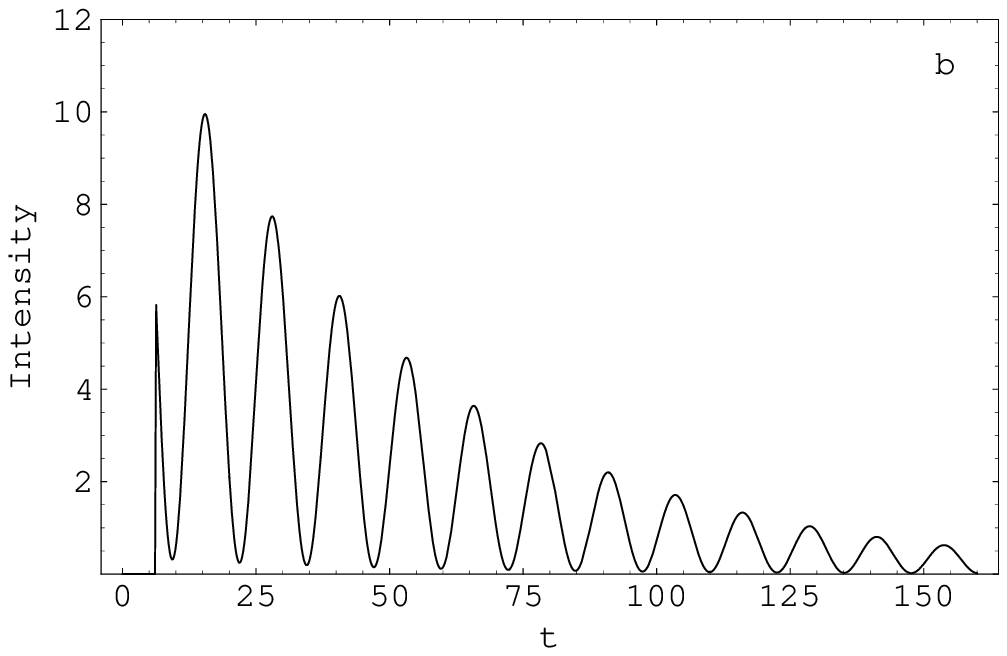}\\[0pt]
\epsfxsize=7.0cm\epsffile{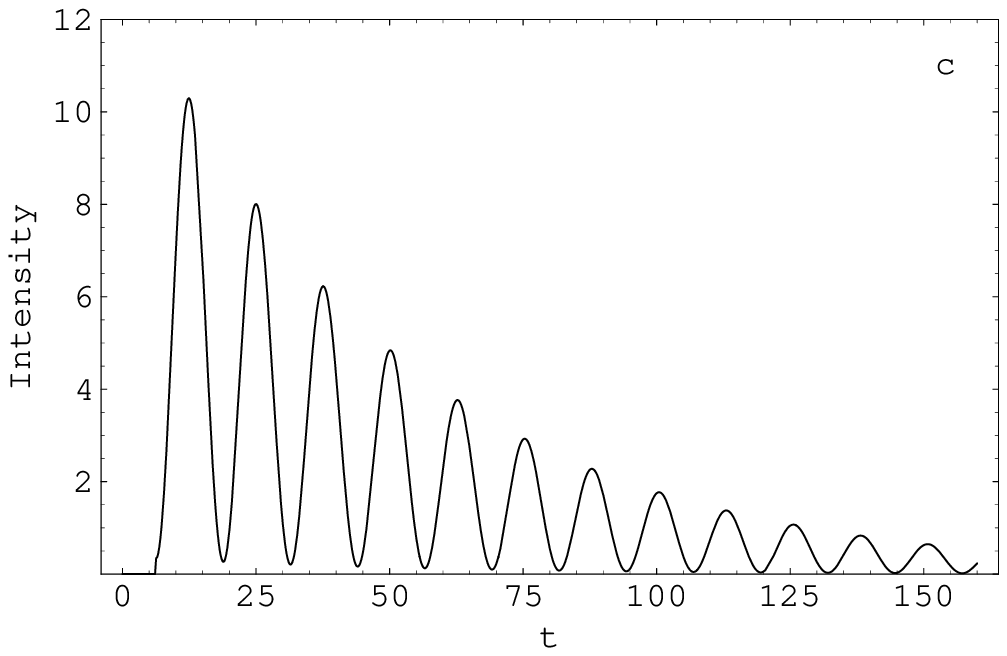}\\[0pt]
\end{center}
\caption{Time evolution of light intensity for the case that the
excitons are initially in a factorized coherent state:
$\protect\beta=\protect\alpha e^{i\protect\phi}$, with
$\protect\phi$ being the phase difference between the two coherent
states. (a) $\phi=0$; (b) $\phi=\pi/2$; (c) $\phi=\pi$; Other
parameters are the same with Fig. 2.}
\end{figure}

The spatial distribution of the light intensity is plotted in Fig.
7 for the case that the two-mode excitons are initially in a
factorized Fock state. Our results show that $I(z,t_0)$ increases
exponentially (see Eq. (\ref{eq:intenFock})) within the regions
$0<|z|<ct_0$ for the case of rotating-wave approximation, and
vanishes immediately as $|z|$ goes beyond $ct_0$. The solid lines
in Fig. 7 are obtained without RWA and shows small-amplitude
oscillations due to the contribution of counter-rotating terms.
Compared with the results of Ref. \cite{Cao}, our results show
that not only the electric field $E(z,t)$ but also the light
intensity $I(z,t)$ do meet the requirement of causal rule.
\vskip 0.1cm
\begin{figure}[hbtp]
\begin{center}
\epsfxsize=7.0cm\epsffile{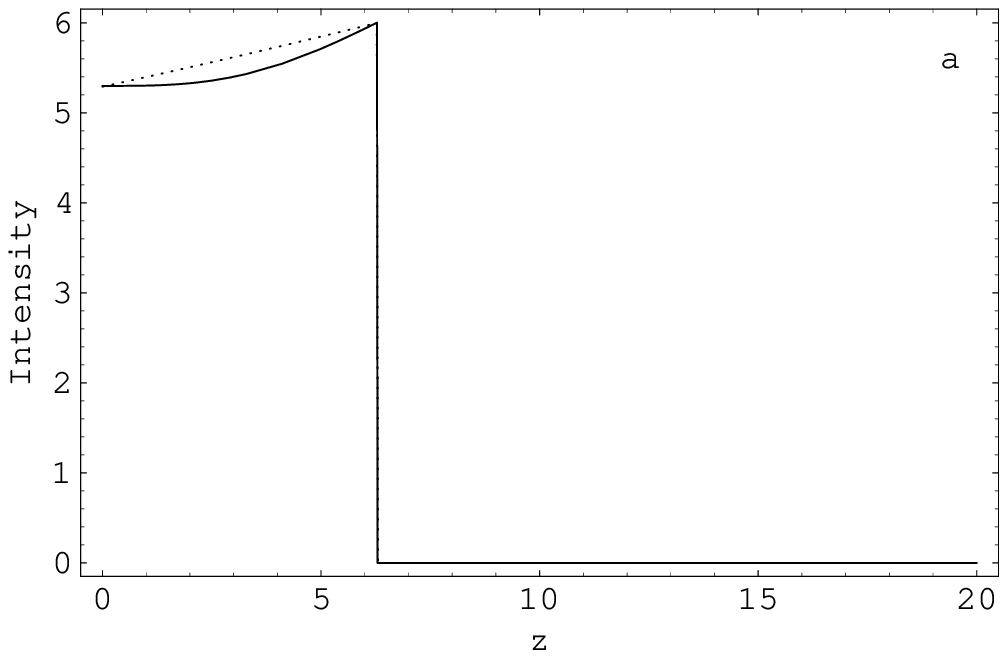}\\[0pt]
\epsfxsize=7.0cm\epsffile{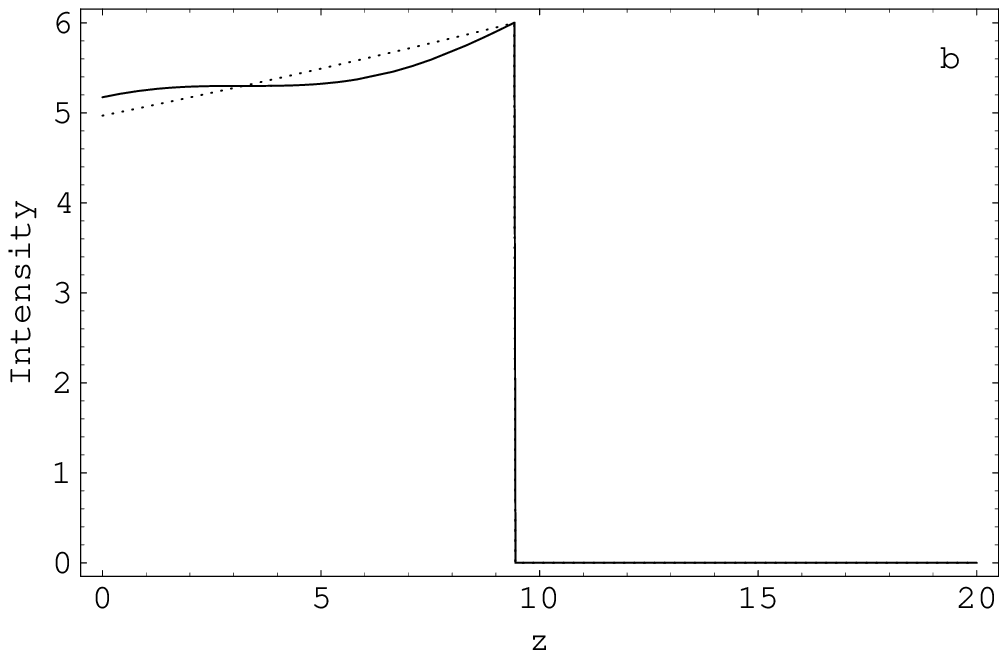}\\[0pt]
\epsfxsize=7.0cm\epsffile{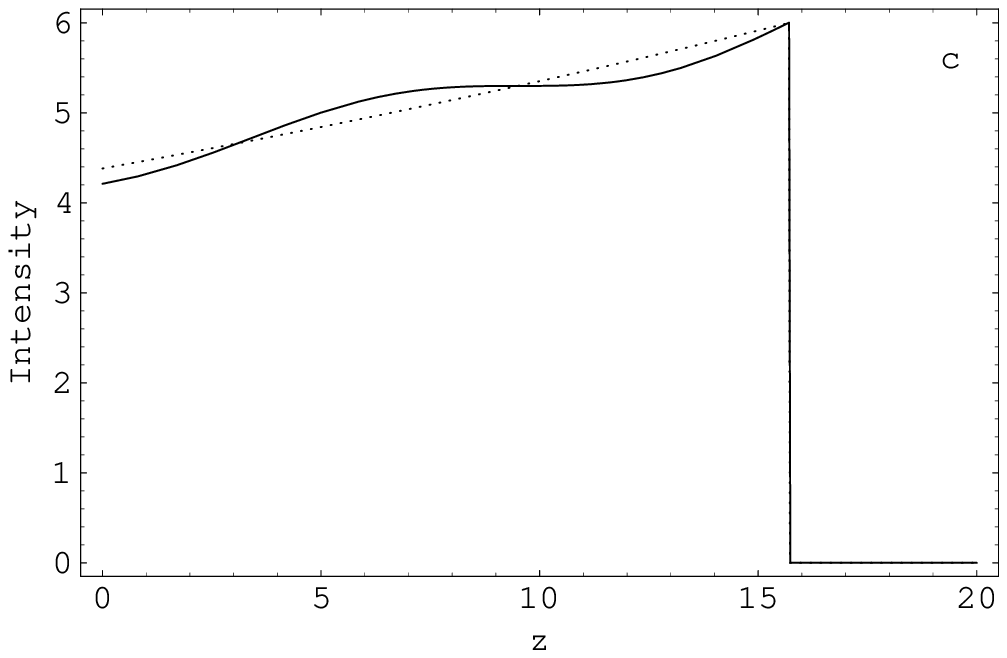}
\end{center}
\caption{Space distribution of intensity $I(z,t_0)$ with (a)$t_0=2
\protect\pi/\Omega_1$, (b)$t_0=3\protect\pi/\Omega_1$, and
(c)$t_0=5\protect \pi/\Omega_1$. $\langle m\rangle _B=\langle
n\rangle _A$, $z-ct$ is in units of $c/\Omega _1$. Other
parameters are the same with Fig. 2. The dot line is obtained by
using RWA, and the solid line is obtained without RWA. }
\end{figure}

\section{conclusion}
In summary, we have studied the collective radiations of a
collection of many V-type three-level atoms in a crystal slab. By
introducing two-mode exciton operators in the large $N$ limits of
the collective quasi-spin operators, these coherent radiations can
be depicted as the fluorescence of low density Frenkel excitons.
The exciton fluorescence exhibits the stronger coherence natures
that the statistical characters of spectrum are identical from the
initial to final stages. This is indeed different from the
ensemble situation with free-moving atoms that the atoms need a
finitely-long time to produce a cooperative radiation, the
enhanced fluorescence.

As a main result of this paper, the occurrence of the quantum beat
aroused from which quantum states of Frenkel exciton is
investigated in this paper. Our results show that not all the
quantum states of excitons may lead to the oscillating behavior in
the light field radiated from the two-mode excitons. This quantum
interference phenomenon may be observed when the two-mode excitons
are initially in a factorized coherent state or an entangled
state. We expect that our theoretical study of quantum beat would
be helpful in practical experiment to measure quantum states of
excitons. Further study of the anharmonic exciton-exciton
interaction in the model is needed. It is also pointed out that,
the algebraic consideration for the definition of the multi-mode
excitons can be generalized to study other exciton system, e.g.,
the quasi-spin wave collective excitations of $\Lambda$-type atom
collection in lattice of crystal that can be used as a new type
quantum memory.

\acknowledgments This work is supported by the NSF of China and
the knowledged Innovation Program(KIP) of the Chinese Academy of
Science. It is also founded by the National Fundamental Research
Program of China with No 001GB309310. Y. X. Liu is supported by
Japan Society for the Promotion of Science (JSPS). The authors
would like to thank C. Q. Cao, W. M. Liu, D. L. Zhou and S. X. Yu
for valuable discussions.

\appendix

\section{SU(3) algebra structure of exciton operators for many-atom system}
From a point of view based on the representation of Lie algebra,
this section describes the mathematical origin of definition of
exciton operator. Physically, this description will clarify why
the conception of the exciton based on the collective operators of
many atoms can be valid only in the case of low excitation. We
first discuss SU(2) algebra structure of excitonic operators for
the two-level many-atom system since the SU(3) case for the 2D
crystal slab containing V-type three-level atoms shares the same
basic idea as that of SU(2). The detailed discussions for the
SU(3) case will follow that of SU(2) in this appendix.

We consider an ensemble of $N$ two-level atoms with their ground
states $|g\rangle^j$ and the excited ones $|e\rangle^j$. Since we
can define the quasi-spin with the Pauli operators
\begin{equation}
\sigma_{-}^{i}=(\sigma_{+}^{i})^{\dag}=|g\rangle^{ii} \langle
e|\text{, } \sigma_{z}^{i}=|e\rangle^{ii} \langle
e|-|g\rangle^{ii} \langle g|,
\end{equation}
the total quasi-angular momentum operators
\begin{equation}
\hat{J}_{-}=\sum_{i=1}^{N}\sigma_{-}^{i},
\hat{J}_{+}=(\hat{J}_{-})^{\dag},
\hat{J}_{z}=\frac{1}{2}\sum_{i=1}^{N}\sigma_{z}^{i},
\end{equation}
define a representation with the highest weight $J=N/2$. The
$(2J+1)$-dimensional irreducible spinor representation of SU(2) in
the symmetric subspace is embedded in the total Hilbert space of
dimension $2^N$.

For a physics system with dynamical SU(2) symmetry, its
Hamiltonian $\hat{H}=\hat{H}(\hat{J}_{-},\hat{J}_{+},\hat{J}_{z})$
is a functional of $\hat{J}_{\pm}$ and $\hat{J}_{z}$. Since the
Casimir operator $\hat{J}^2$ commutes with $\hat{J}_{\pm}$ and
$\hat{J}_{z}$, the eigenvalue $J(J+1)$ will keep conservation in
the time evolution. Here, $J$ can take one of the integers and
half-integers $\frac{N}{2}, \frac{N}{2}-1, ..., 0$. In a physical
process, which one of these $J$ takes depends on the initial state
of the atomic ensemble. The symmetric state $|J=\frac{N}{2},
M=-\frac{N}{2}\rangle=\prod_{i=1}^{N}|g\rangle^i$ represents the
``condensate" with all atoms filling in the ground state. It is
very similar to the situation of an electronic system that the
filling in ground state forms the Fermi surface. In this sense, we
can introduce the atomic collective excitation operators
\begin{equation}
\hat{B}^{\dagger}=\frac{1}{\sqrt{N}} \hat{J}_{+},
\hat{B}=\frac{1}{\sqrt{N}} \hat{J}_{-},
\end{equation}
which is very similar to the exciton operators for an
electron-hole pair. Considering that
\begin{equation}
\hat{J}_{z}=\frac{1}{2}\left[1-\sqrt{(N+1)^{2}-4\hat{J}_{+}\hat{J}_{-}}\right],
\end{equation}
as one solution for the basic angular momentum relation
\begin{eqnarray}
&&\hat{J}_{x}^{2}+\hat{J}_{y}^{2}+\hat{J}_{z}^{2}=\frac{N}{2}\left(\frac{N}{2}+1\right),\\
&&\hat{J}_{+}\hat{J}_{-}=\hat{J}_{x}^{2}+\hat{J}_{y}^{2}+\hat{J}_{z}=\frac{N}{2}
\left(\frac{N}{2}+1\right)-\hat{J}_{z}^{2}+\hat{J}_{z},
\end{eqnarray}
the commutation relation for atomic collective excitation
operators
\begin{equation}
\left[\hat{B}\text{,
}\hat{B}^{\dagger}\right]=-\frac{2}{N}\hat{J}_{z},
\end{equation}
will be reduced to the bosonic relation $\left[\hat{B},
\hat{B}^{\dagger}\right]=1$ for those angular momentum state
$|J,M\rangle$ with much smaller $M$ in comparison to $N=2J$. That
is to say, in the limit with very large $N$ and low excitation,
the collective excitation behave as a boson and thus we call it
atomic exciton. The detailed proof for this was given in ref.
\cite{Liu1} by considering the physical realization of q-deformed
boson algebra \cite{sun-fu} for a very large, but finite $N$. The
main point to prove that is to consider that
\begin{equation}
\frac{\hat{J}_{z}}{N}=\frac{1}{2}\left[\frac{1}{N}-
\sqrt{\left(\frac{1}{N}+1\right)^{2}-\frac{4}{N^2}\hat{J}_{+}\hat{J}_{-}}\right],
\end{equation}
approaches $-1/2$ for the infinite $N$ and the low excitation.
Since the operators can make sense by acting on the symmetric
space, only for those low excitation states $|J,M\rangle$ with
very small $M$ can let the value of $\hat{J}_{+}\hat{J}_{-}/N^2$
approach zero, so that the bosonic commutation relation is
obtained. We can also prove that
\begin{eqnarray}
\hat{B}^{\dagger}\hat{B} &=&\sum_{i=1}^{N}|e\rangle ^{ii} \langle
e|+O\left(\frac{1}{N}\right),
\end{eqnarray}
which means that the free part of the many-atom Hamiltonian can
rationally be described as a free boson in the large $N$ and the
low excitation.

Because the ``condensate" in a ground state plays the crucial role
in defining the exciton operators, the introduction of exciton
operators has to depend on the configuration of the atoms and thus
on the form of interaction. This is just the line, along which we
will define the exciton operators for the V-type atomic system
with the interaction Hamiltonian\cite{Sun2}
\begin{eqnarray}
\hat{H}_I&=&g_{1}\hat{a}^{\dag} \sum_{i=1}^{N}|g\rangle^{ii}
\langle e_{1}|+g_{2}\hat{a}^{\dag} \sum_{i=1}^{N}|g\rangle^{ii}
\langle e_{2}|+H.c..
\end{eqnarray}

It is easy to prove that
\begin{eqnarray}
&&E_1=\sum_{i=1}^{N}|e_{1}\rangle^{ii} \langle g|,
F_1=\sum_{i=1}^{N}|g\rangle^{ii} \langle e_{1}|, \nonumber \\
&&E_2=\sum_{i=1}^{N}|g\rangle^{ii} \langle e_{2}|,
F_2=\sum_{i=1}^{N}|e_{2}\rangle^{ii} \langle g|, \nonumber \\
&&H_1=\frac{1}{2}\sum_{i=1}^{N}\left(|e_{1}\rangle^{ii} \langle e_1|
-|g\rangle^{ii} \langle g|\right),\nonumber \\
&&H_2=\frac{1}{2}\sum_{i=1}^{N}\left(|g\rangle^{ii} \langle g|-
|e_{2}\rangle^{ii} \langle e_2|\right),
\end{eqnarray}
generate a SU(3) algebra with the Cartan subalgebra spanned by
$H_1$ and $H_2$. The basic commutation relations are
\begin{eqnarray}
\left[H_1,E_1\right]&=&E_1,
\left[H_1,F_1\right]=-F_1, \nonumber \\
\left[H_2,E_2\right]&=&E_2,
\left[H_2,F_2\right]=-F_2, \nonumber \\
\left[H_2,E_1\right]&=&-\frac{E_1}{2},
\left[ H_2,F_1\right]=\frac{F_1}{2},\nonumber \\
\left[H_1,E_2\right]&=&-\frac{E_2}{2},
\left[H_1,F_2\right]=\frac{F_2}{2}.
\end{eqnarray}
Two sets of the operators \{$E_1$, $F_1$, $H_1$\} and \{$E_2$,
$F_2$ ,$H_2$\} generate two non-commutation SU(2) subalgebras,
respectively. Actually, the above four collective operators define
a spinor realization for the symmetric representation of SU(3) in
$N_L+1$ dimensional space. The atom number $N_L$ determines the
dimensions $N_L+1$ of representations. In this sense, we can
understand the two-mode excitation in terms of the large $N_L$
limit of representations of SU(3), which just corresponds to the
low density excitation region. Thus we can define the collective
operators
\begin{eqnarray}
\hat{A}=\frac{1}{{\sqrt{N}}}F_1 \text{, }
\hat{A}^{\dag}=\frac{1}{{\sqrt{N}}}E_1,\nonumber \\
\hat{B}=\frac{1}{{\sqrt{N}}}E_2 \text{, }
\hat{B}^{\dag}=\frac{1}{{\sqrt{N}}}F_2.
\end{eqnarray}
For a very large $N$ and low excitation, it is easy to prove that
$\hat{A}$ and $\hat{B}$ commute with each other and obey the
standard bosonic commutation relation.

Now we can consider the construction of Frenkel excitonic states.
In this way, the initial conditions for Frenkel exciton can be
given in terms of the single atom preparations. For example, when
\text{$n=0$, $1$, $2$, ...}, the Fock states of the $A$-mode
excitons
\begin{equation}
|n\rangle_A =\frac{1}{\sqrt{n!}}\left(\hat{A}^{\dag}\right)^
n|0\rangle =\frac{1}{N^{n/2}
\sqrt{n!}}\left(\sum_{i=1}^{N}|e_{1}\rangle ^{ii}\langle
g|\right)^{n}|0\rangle
\end{equation}
take the symmetric excitation states
\begin{eqnarray}
&&|0\rangle=|g,g,...g\rangle,\nonumber \\
&&|1\rangle_A
=\frac{1}{\sqrt{N}}\sum_{j=1}^{N}|g,g,...,e^{j}_1,...,g
\rangle , \nonumber \\
&&|2\rangle_A =\frac{1}{N\sqrt{2}}\sum_{j,
k=1}^{N}|g,g,...,e^{j}_1,..,e^{k}_1,...,g\rangle, \nonumber \\
&&....
\end{eqnarray}
The second example is the coherent state of the $A$-mode Frenkel
exciton
\begin{eqnarray}
|\alpha\rangle_A &\propto& \exp(\alpha \hat{A}^{\dagger
})|0\rangle\nonumber \\
&=& \prod_{j=1}\left[\cos \theta |g\rangle ^{j}+\sin \theta
e^{i\phi }|e_1\rangle ^{j}\right],
\end{eqnarray}
where $\tan \theta =\frac{|\alpha |}{\sqrt{N}}$ and $\alpha
=|\alpha |e^{i\phi }$. The coherent nature of this many-atomic
state is reflected by the fact that both $\theta $ and $\phi $ are
independent of the index $j$ of atoms. The quantum states of
$B$-mode excitons can also be constructed by using the same
procedure as discussed above.

\section{The self-interaction term of the light field}

In this section we will calculate ${\bf e}_{q\lambda }\cdot {\bf
e}_{q^{\prime }\lambda ^{\prime }}$ for the three-level case. For
an atom with a complete set of eigenvectors $\{|n\rangle\}$, we
have
\begin{eqnarray}
{\bf e}_{q\lambda }\cdot {\bf e}_{q^{\prime }\lambda ^{\prime }}
&=&\langle g|{\bf e}_{q\lambda }\cdot {\bf e}_{q^{\prime }\lambda
^{\prime }}|g\rangle
\nonumber \\
&=&\frac 1{i\hbar }\sum_n\langle g|{\bf e}_{q\lambda }\cdot {\bf
x|}n\rangle \langle n|{\bf p\cdot e}_{q^{\prime }\lambda ^{\prime
}}|g\rangle  \nonumber
\\
&&-\sum_n\langle g|{\bf e}_{q\lambda }\cdot {\bf p|}n\rangle
\langle n|{\bf
x\cdot e}_{q^{\prime }\lambda ^{\prime }}|g\rangle  \nonumber \\
&=&\frac{2m}{\hbar e^2}\sum_n\Omega _n\langle g|{\bf e}_{q\lambda
}\cdot {\bf d|}n\rangle \langle n|{\bf d\cdot e}_{q^{\prime
}\lambda ^{\prime }}|g\rangle ,
\end{eqnarray}
where we have used $\left[ {\bf x},{\bf p}\right] =i\hbar $ and
${\bf p}= \frac m{i\hbar }\left( {\bf x}\hat{H}_A-\hat{H}_A{\bf
x}\right)$. $\hat{H}_A$ is the free atomic Hamiltonian and gives
the eigenvalue equation: $\hat{H}_A|n\rangle =E_n |n\rangle $.

For the V-type three-level case, we take the three-level
approximation in the above equation as ref. \cite{Cao} for the two
level atom case, obtaining
\begin{eqnarray}
{\bf e}_{q\lambda }\cdot {\bf e}_{q^{\prime }\lambda ^{\prime }}
&\approx&\frac{2m}{\hbar e^2}\left[\Omega _1\left( {\bf
e}_{q\lambda } \cdot {\bf d}_1\right)\left( {\bf d}_1{\bf \cdot
e}_{q^{\prime }
\lambda ^{\prime }}\right)\right.\nonumber \\
&&\left.+\Omega _2\left( {\bf e}_{q\lambda }\cdot {\bf d}_2\right)
 \left( {\bf d}_2{\bf \cdot e}_{q^{\prime }\lambda ^{\prime }}\right) \right],
\end{eqnarray}
where $\Omega _1=(E_{e_1}-E_g)/\hbar$ and $\Omega
_2=(E_{e_2}-E_g)/\hbar$ are atomic transition frequencies for
$\left| g\right\rangle \leftrightarrow \left| e_1\right\rangle $
and $\left| g\right\rangle \leftrightarrow \left| e_2\right\rangle
$, respectively. ${\bf d}_1=\langle e_1| {\bf d}|g\rangle =\langle
g|{\bf d}|e_1\rangle $ and ${\bf d}_2=\langle e_2| {\bf
d}|g\rangle =\langle g|{\bf d}|e_2\rangle$ are the corresponding
transition dipole moments. Choosing ${\bf e}_{q\lambda }$ as the
following
\begin{eqnarray}
{\bf e}_{q1}\cdot {\bf d}_1 &=&d_1\text{, }{\bf e}_{q1}\cdot {\bf
d}_2=d_2, \nonumber \\
{\bf e}_{q^{\prime }2}\cdot {\bf d}_1 &=&0\text{, }{\bf
e}_{q^{\prime }2}\cdot {\bf d }_2=0,
\end{eqnarray}
then we get
\begin{equation}
{\bf e}_{q\lambda }\cdot {\bf e}_{q^{\prime }\lambda ^{\prime
}}\approx\frac{2m}{\hbar e^2}[\Omega _1d_1^2\delta _{\lambda
,1}\delta _{\lambda ,\lambda ^{\prime }}+\Omega _2d_2^2\delta
_{\lambda ,1}\delta _{\lambda ,\lambda ^{\prime }}].
\end{equation}
Substituting $\sum_{\lambda,\lambda ^{\prime
}}\mathbf{e}_{q\lambda }\cdot \mathbf{e}_{q^{\prime }\lambda
^{\prime }}$ into $f(q,q^{\prime })$, we get Eq. (45) in section
V. It is noticed that, for the case with single direction
polarization of light, $\sum_{\lambda ,\lambda ^{\prime
}}\mathbf{e}_{q\lambda }\cdot \mathbf{e}_{q^{\prime }\lambda
^{\prime }}=1$ strictly. The cut-off of the complete relation for
the sum by only three levels, however, will lead to the departure
from $1$. Only with this cut-off approximation the
exactly-solvable mode is built for the two mode excitons coupling
to the quantized electromagnetic fields.

\section{Detailed non-perturbation calculation}
In this Appendix, we will add the necessary details and list the
more expatiatory expressions for section V. We start from the
total interaction Hamiltonian
\begin{eqnarray}
\hat{H}_{int}&=&\hbar\sum_{q,k}G_1(q)O(k+q)\left[\hat{A}_k+\hat{A}_{-k}^{\dagger
}\right]\left[\hat{a}_q+\hat{a}_{-q}^{\dagger }\right]  \nonumber \\
&+&\hbar\sum_{q,k}G_2(q)O(k+q)\left[\hat{B}_k+\hat{B}_{-k}^{\dagger
}\right] \left[\hat{a}_q+\hat{a}_{-q}^{\dagger }\right]  \nonumber \\
&+&\hbar \sum_{q,q^{\prime },k}\frac 1{\Omega
_1}G_1(q)G_1(q^{\prime })O(q^{\prime }-k)O(k+q)\nonumber \\
&&\times\left[ \hat{a}_q+\hat{a}_{-q}^{\dagger }\right]
\left[\hat{a}_{q^{\prime}}+\hat{a}_{-q^{\prime }}^{\dagger }\right]
\nonumber \\
&+&\hbar \sum_{q,q^{\prime },k}\frac 1{\Omega
_2}G_2(q)G_2(q^{\prime })O(q^{\prime }-k)O(k+q)\nonumber \\
&&\times \left[ \hat{a}_q+\hat{a}_{-q}^{\dagger }\right]
\left[\hat{a}_{q^{\prime }}+\hat{a}_{-q^{\prime }}^{\dagger
}\right].
\end{eqnarray}
which includes the non-RWA terms and the self-interaction of the
light field.

The half-side Fourier transformation (HSFT) of the Heisenberg
equations governed by the total Hamiltonian (C1) are
\begin{eqnarray}
(\omega -\Omega _1)A_k(\omega ) &=&\sum_qG_1(q)O(q-k)\left[
a_q(\omega )+a_{-q}^{\dagger }(\omega )\right]\nonumber \\
&&+iA_k(0), \\
(\omega -\Omega _2)B_k(\omega ) &=&\sum_qG_2(q)O(q-k)\left[
a_q(\omega )+a_{-q}^{\dagger }(\omega )\right]\nonumber \\
&&+iB_k(0),
\end{eqnarray}
for excitons, and
\begin{eqnarray}
&&(\omega -|q|c)a_q(\omega ) \nonumber\\
&=&\frac \omega {\Omega _1}G_1(q)\sum_kO(k-q)\left[ A_k(\omega
)-A_{-k}^{\dagger }(\omega )\right] \nonumber\\
&-&i\frac 1{\Omega _1}G_1(q)\sum_kO(k-q)\left[
A_k(0)-A_{-k}^{\dagger}(0)\right]\nonumber \\
&+&\frac \omega {\Omega _2}G_2(q)\sum_kO(k-q)\left[ B_k(\omega
)-B_{-k}^{\dagger }(\omega )\right] \nonumber\\
&-&i\frac 1{\Omega _2}G_2(q)\sum_kO(k-q)\left[
B_k(0)-B_{-k}^{\dagger }(0)\right]\nonumber\\
&+&ia_q(0),\label{eq:aq}
\end{eqnarray}
\begin{eqnarray}
&&(\omega +|q|c)a_{-q}^{\dagger }(\omega )  \nonumber \\
&=&-\frac \omega {\Omega _1}G_1(q)\sum_kO(k-q)\left[
A_k(\omega )-A_{-k}^{\dagger }(\omega )\right] \nonumber \\
&+&i\frac 1{\Omega _1}G_1(q)\sum_kO(k-q)\left[
A_k(0)-A_{-k}^{\dagger}(0)\right]  \nonumber \\
&-&\frac \omega {\Omega _2}G_2(q)\sum_kO(k-q)\left[ B_k(\omega
)-B_{-k}^{\dagger }(\omega )\right]  \nonumber \\
&+&i\frac 1{\Omega _2}G_2(q)\sum_kO(k-q)\left[
B_k(0)-B_{-k}^{\dagger }(0)\right]\nonumber \\
&+&ia_{-q}^{\dagger }(0) ,
\end{eqnarray}
for photons. Here, we have eliminated $\sum_{q^{\prime
}}G(q^{\prime })O(q^{\prime }-k)\left[ a_{q^{\prime }}(\omega
)+a_{-q^{\prime }}^{\dagger }(\omega )\right]$ in the derivation
of Eqs. (C4) and (C5).

Combining the above two equations, we get
\begin{eqnarray}
&&(\omega ^2-q^2c^2)\left[ a_q(\omega )+a_{-q}^{\dagger }(\omega
)\right]\nonumber \\
&=&i(\omega +|q|c)a_q(0)+i(\omega -|q|c)a_{-q}^{\dagger }(0)  \nonumber \\
&&+2|q|c\frac \omega {\Omega _1}G_1(q)\sum_kO(k-q)\left[
A_k(\omega)-A_{-k}^{\dagger }(\omega )\right]  \nonumber \\
&&-i\frac{2|q|c}{\Omega _1}G_1(q)\sum_kO(k-q)\left[
A_k(0)-A_{-k}^{\dagger}(0)\right]  \nonumber \\
&&+2|q|c\frac \omega {\Omega _2}G_2(q)\sum_kO(k-q)\left[
B_k(\omega
)-B_{-k}^{\dagger }(\omega )\right]  \nonumber \\
&&-i\frac{2|q|c}{\Omega _2}G_2(q)\sum_kO(k-q)\left[
B_k(0)-B_{-k}^{\dagger }(0)\right].
\end{eqnarray}
Substituting Eq. (C6) into Eq. (C2) and (C3), after a
straightforward calculation, we obtain two coupled equations for
the Frenkel exciton operators
\begin{eqnarray}
&&\sum_{k^{\prime }}\left[ (\omega ^2-\Omega _1^2)\delta
_{kk^{\prime }}- \frac{2\omega ^2}{\Omega _1}F_{kk^{\prime
}}^{(1)}(\omega )\right]\nonumber\\
&&\times\left[A_{k^{\prime }}(\omega )-A_{-k^{\prime}}^{\dagger
}(\omega )\right]\nonumber\\
&=&i\left[ (\omega +\Omega _1)A_k(0)-(\omega -\Omega
_1)A_{-k}^{\dagger }(0)\right]\nonumber\\
&+&2i\omega \sum_qG_1(q) O(q-k)\left[ \frac{a_q(0)}{\omega
-|q|c}+\frac{a_{-q}^{\dagger }(0) }{\omega +|q|c}\right]
\nonumber\\
&-&2i\frac \omega {\Omega _1}\sum_{k^{\prime }}F_{kk^{\prime
}}^{(1)}(\omega ) \left[ A_{k^{\prime }}(0)-A_{-k^{\prime
}}^{\dagger }(0)\right]\nonumber\\
&-&2i \frac \omega {\Omega _2}\sum_{k^{\prime }}F_{kk^{\prime
}}^{(3)}(\omega )\left[ B_{k^{\prime }}(0)-B_{-k^{\prime
}}^{\dagger }(0)\right]\nonumber\\
&+&\frac{2\omega ^2}{\Omega _2}\sum_{k^{\prime }}F_{kk^{\prime
}}^{(3)}(\omega )\left[ B_{k^{\prime }}(\omega )-B_{-k^{\prime
}}^{\dagger }(\omega )\right] ,
\end{eqnarray}
and
\begin{eqnarray}
&&\sum_{k^{\prime }}\left[ (\omega ^2-\Omega _2^2)\delta
_{k,k^{\prime }}- \frac{2\omega ^2}{\Omega _2}F_{kk^{\prime
}}^{(2)}(\omega )\right]\nonumber \\
&&\times\left[ B_{k^{\prime }}(\omega )-B_{-k^{\prime }}^{\dagger
}(\omega )\right]
\nonumber \\
&=&i\left[ (\omega +\Omega _2)B_k(0)-(\omega -\Omega
_2)B_{-k}^{\dagger }(0)\right]\nonumber\\
&+&2i\omega \sum_qG_2(q)O(q-k)\left[ \frac{a_q(0)}{\omega
-|q|c}+\frac{a_{-q}^{\dagger }(0) }{\omega +|q|c}\right]
\nonumber\\
&-&2i\frac \omega {\Omega _2}\sum_{k^{\prime }}F_{kk^{\prime
}}^{(2)}(\omega ) \left[ B_{k^{\prime }}(0)-B_{-k^{\prime
}}^{\dagger }(0)\right]
\nonumber\\
&-&2i \frac \omega {\Omega _1}\sum_{k^{\prime }}F_{kk^{\prime
}}^{(3)}(\omega )\left[ A_{k^{\prime }}(0)-A_{-k^{\prime
}}^{\dagger }(0)\right]
\nonumber\\
&+&\frac{2\omega ^2}{\Omega _1}\sum_{k^{\prime }}F_{kk^{\prime
}}^{(3)}(\omega )\left[ A_{k^{\prime }}(\omega )-A_{-k^{\prime
}}^{\dagger }(\omega )\right] ,
\end{eqnarray}
where only the initial photon operators and the exciton operators
are concerned. Three factors introduced in Eqs. (c7) and (c8) are
\begin{eqnarray}
F_{kk^{\prime }}^{(1)}(\omega )
&=&\sum_q\frac{2|q|cG_1^2(q)O(q-k)O(k^{
\prime }-q)}{\omega ^2-q^2c^2},  \nonumber \\
F_{kk^{\prime }}^{(2)}(\omega )
&=&\sum_q\frac{2|q|cG_2^2(q)O(q-k)O(k^{
\prime }-q)}{\omega ^2-q^2c^2},  \nonumber \\
F_{kk^{\prime }}^{(3)}(\omega )
&=&\sum_q\frac{2|q|cG_1(q)G_2(q)O(q-k)O(k^{ \prime }-q)}{\omega
^2-q^2c^2},
\end{eqnarray}
which represent the overlap of exciton wave functions with
different wave vectors. We take the photon normalization volume
$V$ to be $AL$ where $A$ is the area of the crystal slab, and
place the slab at the middle of the volume. When $L$ is sufficient
large, the sum over $q$ can be replaced to a integral:
$\sum_q...\rightarrow \frac L{2\pi }\int_{-\infty }^\infty dq...$.
Thus
\begin{eqnarray}
F_{kk^{\prime }}^{(i)}(\omega ) &=&-\frac{Na\Omega _if_i^2}{4\pi
c^2} \int_{-\infty }^\infty dq\frac{O(q-k)O(k^{\prime
}-q)}{q^2-(\frac \omega c)^2 },
\end{eqnarray}

By carrying out the integrations as in Ref. \cite{Cao}, the
explicit expressions are
\begin{eqnarray}
F_{kk^{\prime }}^{(i)}(\omega ) &=&-\frac{af_i^2\Omega
_i}{8Nc\omega }\left[ \frac{\sin \frac{k^{\prime }+\frac \omega
c}2Na}{\sin \frac{k^{\prime }+ \frac \omega
c}2a}\frac{e^{i\frac{k+\frac \omega c}2Na}}{\sin \frac{k+\frac
\omega c}2a}\right.\nonumber\\
&-&\left.\frac{\sin \frac{k^{\prime }-\frac \omega c}2Na}{\sin
\frac{ k^{\prime }-\frac \omega c}2a}\frac{e^{i\frac{k-\frac
\omega c}2Na}}{\sin\frac{k-\frac \omega c}2a}\right.\nonumber \\
&+&\left.\frac{\sin \frac{k-k^{\prime }}2Na}{\sin
\frac{k-k^{\prime }}2a}\frac{ \sin \frac \omega ca}{\sin
(\frac{k+\frac \omega c}2a)\sin (\frac{k-\frac \omega
c}2a)}\right].
\end{eqnarray}
When $N\rightarrow \infty$, the first two terms tend to zero, thus
\begin{equation}
\lim_{N\rightarrow \infty }F_{kk^{\prime }}^{(i)}(\omega )=-\frac{
af_i^2\Omega _i}{8c\omega }\frac{\sin \frac \omega ca}{\sin
(\frac{k+\frac \omega c}2a)\sin (\frac{k-\frac \omega c}2a)}\delta
_{k,k^{\prime }}.
\end{equation}
The above results will be used to determine $a_q(\omega
)-a_{-q}^{\dagger }(\omega )$ explicitly in Sec. V.

For the single lattice layer case, ignoring the quantum noise
terms proportional to $a_q(0)$ or $a_{-q}^{\dagger }(0)$ induced
by the background light field, we obtain the coupled equations of
the exciton operators
\begin{eqnarray}
&&\left[\omega ^2-\Omega _2^2-\frac{2\omega ^2}{\Omega _2}
F_{00}^{(2)}(\omega )\right] \left[ B_0(\omega )-B_0^{\dagger
}(\omega)\right] \nonumber\\
&=&i\left[ (\omega +\Omega _2)B_0(0)-(\omega -\Omega
_2)B_0^{\dagger}(0)\right] \nonumber\\
&&-2i\frac \omega {\Omega _2}F_{00}^{(2)}(\omega )\left[
B_0(0)-B_0^{\dagger}(0)\right] \nonumber\\
&&-2i\frac \omega {\Omega _1}F_{00}^{(3)}(\omega )\left[
A_0(0)-A_0^{\dagger}(0)\right] \nonumber\\
&&+\frac{2\omega ^2}{\Omega _1}F_{00}^{(3)}(\omega )\left[
A_0(\omega )-A_0^{\dagger }(\omega )\right] ,\label{eq:CoupEq1}
\end{eqnarray}
and
\begin{eqnarray}
&&\left[\omega ^2-\Omega _1^2-\frac{2\omega ^2}{\Omega _1}
F_{00}^{(1)}(\omega )\right] \left[ A_0(\omega )-A_0^{\dagger
}(\omega)\right]  \nonumber \\
&=&i\left[ (\omega +\Omega _1)A_0(0)-(\omega -\Omega
_1)A_0^{\dagger}(0)\right]  \nonumber \\
&&-2i\frac \omega {\Omega _1}F_{00}^{(1)}(\omega )\left[
A_0(0)-A_0^{\dagger}(0)\right]  \nonumber \\
&&-2i\frac \omega {\Omega _2}F_{00}^{(3)}(\omega )\left[
B_0(0)-B_0^{\dagger}(0)\right]  \nonumber \\
&&+\frac{2\omega ^2}{\Omega _2}F_{00}^{(3)}(\omega )\left[
B_0(\omega )-B_0^{\dagger }(\omega )\right]. \label{eq:CoupEq2}
\end{eqnarray}
These two equations lead to two decoupled equations Eq.
(\ref{eq:CloseEq1}) and Eq. (\ref{eq:CloseEq2}) in Sec. V for
$A_0(\omega )-A_0^{\dagger }(\omega )$ and $B_0(\omega
)-B_0^{\dagger }(\omega )$. The solutions of Eq.
(\ref{eq:CloseEq1}) and Eq. (\ref{eq:CloseEq2}) are
\begin{eqnarray}
&&A_0(\omega )-A_0^{\dagger }(\omega ) \nonumber\\
&=&i\frac{\omega ^2-\Omega _2^2+i\eta _2\omega }{\zeta (\omega
)}\left[(\omega +\Omega _1)A_0(0)-(\omega -\Omega _1)A_0^{\dagger
}(0)\right]\nonumber\\
&&+\frac{\Omega _3\eta _3}{\zeta (\omega )}\left[ (\omega +\Omega
_2)B_0(0)-(\omega -\Omega _2)B_0^{\dagger }(0)\right] \nonumber\\
&&-\frac{\eta _1}{\zeta (\omega )}\left( \omega ^2-\Omega
_2^2\right) \left[ A_0(0)-A_0^{\dagger }(0)\right]
,\label{eq:sol1}
\end{eqnarray}
and
\begin{eqnarray}
&&B_0(\omega )-B_0^{\dagger }(\omega )  \nonumber \\
&=&i\frac{\omega ^2-\Omega _1^2+i\eta _1\omega }{\zeta (\omega
)}\left[ (\omega +\Omega _2)B_0(0)-(\omega -\Omega _2)B_0^{\dagger
}(0)\right]
\nonumber \\
&&+\frac{\Omega _3\eta _3}{\zeta (\omega )}\left[ (\omega +\Omega
_1)A_0(0)-(\omega -\Omega _1)A_0^{\dagger }(0)\right]  \nonumber \\
&&-\frac{\eta _2}{\zeta (\omega )}\left( \omega ^2-\Omega
_1^2\right) \left[ B_0(0)-B_0^{\dagger }(0)\right],\label{eq:sol2}
\end{eqnarray}
where $\zeta (\omega )$ is defined in Eq. (\ref{eq:zeta}). The
above two equations determine $a_q(\omega )-a_{-q}^{\dagger
}(\omega )$ and give the non-zero electric field $E(z,t)$ (Eq.
(65) in Sec. V)
\begin{eqnarray*}
E(z,t)={\mathcal E^{(+)}}(z,t) +{\rm H.c.},
\end{eqnarray*}
for $z-ct<0$, where
\begin{eqnarray}
&&{\mathcal E^{(+)}}(z,t)\nonumber\\
&=&\sqrt{\frac{\pi \hbar \Omega _{1}\eta _{1}}{cA}}\left[ \left(
\omega _{1}+\Omega _{1}\right) A_{0}(0)+\left( \omega _{1}-\Omega
_{1}\right) A_{0}^{\dagger }(0)\right]   \nonumber \\
&&\times \frac{(\omega _{1}^{2}-\Omega _{2}^{2})e^{-i\omega
_{1}(t-\frac{z}{c })}}{\left( \omega _{1}-\omega _{2}\right)
\left( \omega _{1}-\omega
_{3}\right) \left( \omega _{1}-\omega _{4}\right) }  \nonumber \\
&&+\sqrt{\frac{\pi \hbar \Omega _{1}\eta _{1}}{cA}}\left[ \left(
\omega _{3}+\Omega _{1}\right) A_{0}(0)+\left( \omega _{3}-\Omega
_{1}\right)
A_{0}^{\dagger }(0)\right]   \nonumber \\
&&\times \frac{(\omega _{3}^{2}-\Omega _{2}^{2})e^{-i\omega
_{3}(t-\frac{z}{c })}}{\left( \omega _{3}-\omega _{1}\right)
\left( \omega _{3}-\omega
_{2}\right) \left( \omega _{3}-\omega _{4}\right) }  \nonumber \\
&&+\sqrt{\frac{\pi \hbar \Omega _{2}\eta _{2}}{cA}}\left[ \left(
\omega _{1}+\Omega _{2}\right) B_{0}(0)+\left( \omega _{1}-\Omega
_{2}\right)
B_{0}^{\dagger }(0)\right]   \nonumber \\
&&\times \frac{(\omega _{1}^{2}-\Omega _{1}^{2})e^{-i\omega
_{1}(t-\frac{z}{c })}}{\left( \omega _{1}-\omega _{2}\right)
\left( \omega _{1}-\omega
_{3}\right) \left( \omega _{1}-\omega _{4}\right) }  \nonumber \\
&&+\sqrt{\frac{\pi \hbar \Omega _{2}\eta _{2}}{cA}}\left[ \left(
\omega _{3}+\Omega _{2}\right) B_{0}(0)+\left( \omega _{1}-\Omega
_{2}\right)
B_{0}^{\dagger }(0)\right]   \nonumber \\
&&\times \frac{(\omega _{3}^{2}-\Omega _{1}^{2})e^{-i\omega
_{3}(t-\frac{z}{c })}}{\left( \omega _{3}-\omega _{1}\right)
\left( \omega _{3}-\omega _{2}\right) \left( \omega _{3}-\omega
_{4}\right) },\label{eq:elecfield}
\end{eqnarray}
here $\omega _j$, for $j=1,2,3,4$, are four roots of the
characteristic equation of Eq. (\ref{eq:CharaEq}).  Eq.
(\ref{eq:elecfield}) shows that, in positive $z$ region, the
electric field generated by the two-mode exciton is damped
exponentially with two eigendecay rates $\Gamma _1$ and $\Gamma
_2$. Unlike the results of Sec. IV, the corresponding eigenmodes,
however, are two linear combinations of the $A_0$ and $B_0$ modes.
This is because we include non-RWA terms and MPP in the derivation
of Eq. (\ref{eq:CloseEq1}) and Eq. (\ref{eq:CloseEq2}) in Sec. V.


\end{multicols}
\end{document}